\documentclass[10pt,journal,compsoc]{IEEEtran}
\ifCLASSOPTIONcompsoc
  \usepackage[nocompress]{cite}
\else
  \usepackage{cite}
\fi
\ifCLASSINFOpdf
\else
\fi
\usepackage{amsmath,amssymb,amsfonts}
\usepackage{algorithmic}
\usepackage{graphicx}
\usepackage{textcomp}
\newcommand{\etal}{\textit{et al}.}
\newcommand{\STAB}[1]{\begin{tabular}{@{}c@{}}#1\end{tabular}}
\newcommand{\specialcell}[2][c]{\begin{tabular}[#1]{@{}l@{}}#2\end{tabular}}
\usepackage[shortlabels]{enumitem}
\usepackage{tikz}
\usepackage{xcolor}
\usepackage{multirow}
\usepackage[caption=false,font=normalsize,labelfont=sf,textfont=sf]{subfig}
\usepackage[subpreambles=true]{standalone}
\usepackage{hyperref}
\hypersetup{breaklinks=true}
\newcommand\Tstrut{\rule{0pt}{2.6ex}}       % "top" strut
\newcommand\Bstrut{\rule[-0.9ex]{0pt}{0pt}} % "bottom" strut
\newcommand{\TBstrut}{\Tstrut\Bstrut} % top&bottom struts

\begin{document}
\title{Artificial Emotional Intelligence in Socially Assistive Robots for Older Adults: A Pilot Study}
\author{Hojjat~Abdollahi$^{1,2}$,~\IEEEmembership{Student Member,~IEEE,}
        Mohammad~H.~Mahoor$^{1,2}$,~\IEEEmembership{Senior Member,~IEEE}, \\Rohola Zandie$^{1,2}$,~\IEEEmembership{Student Member,~IEEE,} Jarid Siewierski$^{2}$, and Sara H. Qualls$^{3}$% <-this % stops a space
\thanks{$^{1}$Department of Electrical and Computer Engineering, University of Denver, E-mail: (hojjat.abdollahi, mmahoor, roholazandie)@du.edu $^{2}$DreamFace Technologies, LLC. $^{3}$Department of Psychology and Gerontology Center, University of Colorado, Colorado Springs squalls@uccs.edu}% <-this % stops a space
%\thanks{J. Doe and J. Doe are with Anonymous University.}% <-this % stops a space
%\thanks{Manuscript received April 19, 2005; revised August 26, 2015.}
}
\vskip -20mm

% \markboth{IEEE Transactions on Affective Computing}
% {Abdollahi \MakeLowercase{\textit{et al.}}: Artificial Emotional Intelligence in Socially Assistive Robots for Older Adults: A Pilot Study}

\IEEEtitleabstractindextext{
\begin{abstract}

This paper presents our recent research on integrating artificial emotional intelligence in a social robot (Ryan) and studies the robot's effectiveness in engaging older adults. Ryan is a socially assistive robot designed to provide companionship for older adults with depression and dementia through conversation. We used two versions of Ryan for our study, empathic and non-empathic. The empathic Ryan utilizes a multimodal emotion recognition algorithm and a multimodal emotion expression system. Using different input modalities for emotion, i.e. facial expression and speech sentiment, the empathic Ryan detects users’ emotional state and utilizes an affective dialogue manager to generate a response. On the other hand, the non-empathic Ryan lacks facial expression and uses scripted dialogues that do not factor in the  users’ emotional state. We studied these two versions of Ryan with 10 older adults living in a senior care facility. The statistically significant improvement in the users' reported face-scale mood measurement indicates an overall positive effect from the interaction with both the empathic and non-empathic versions of Ryan. However, the number of spoken words measurement and the exit survey analysis suggest that the users perceive the empathic Ryan as more engaging and likable. 

\end{abstract}

% Note that keywords are not normally used for peerreview papers.
}

\maketitle

\IEEEdisplaynontitleabstractindextext
\IEEEpeerreviewmaketitle
\IEEEraisesectionheading{
\section{Introduction}
\label{sec:introduction}}

Socially Assistive Robotics (SAR) is a sub-field in robotics that aims to develop intelligent robots that can provide aid and support to users~\cite{feil2005defining}. For instance, older adults living in senior care facilities often feel lonely and isolated. Social interaction and mental stimulation are critical for improving their well-being~\cite{banerjee2020social,aung2017loneliness }. SAR has shown to alleviate this problem by providing companionship to assist older adults through conversation and social interaction~\cite{ghafurian2021social, vandemeulebroucke2018older}.
Furthermore, the global outbreak of COVID-19 and the effects of social distancing and stay-at-home orders drew more attention to the isolation of older adults living in senior care facilities. The COVID-19 pandemic has highlighted the healthcare worker shortage that currently plagues the healthcare system~\cite{xu2020shortages}, and SAR has recently been used to address this problem by researchers~\cite{chen2020social, adams_social_2020, henkel2020robotic}.

To more naturally and effectively interact with humans, we can endow robots with social capabilities.  
A social robot must be equipped~\cite{tapus2007grand} with human-oriented interaction that exhibits context and user-appropriate social behavior and focuses attention and communication on the user. Studies suggest that adding emotional information to SAR enhances user satisfaction~\cite{prendinger2005empathic} and results in a more positive interaction between robot and human. Empathy is a critical skill in health and elder care; users perceive robots that express empathic behavior as more friendly, understanding, and caring~\cite{bagheri2021reinforcement}.  

A social robot with Artificial Emotional Intelligence (AEI) can recognize, process, simulate, and react to human affects/emotions~\cite{yonck2020heart}. The development of affective and empathic robots that have the capability to recognize users' emotions and interact with them naturally and effectively is in its infancy and more research needs to be carried out in this field~\cite{pu2019effectiveness}.

\begin{figure}
  \centering
%   \includestandalone[width=.5\textwidth, mode=buildnew]{experiment}
    \includegraphics[scale=.45]{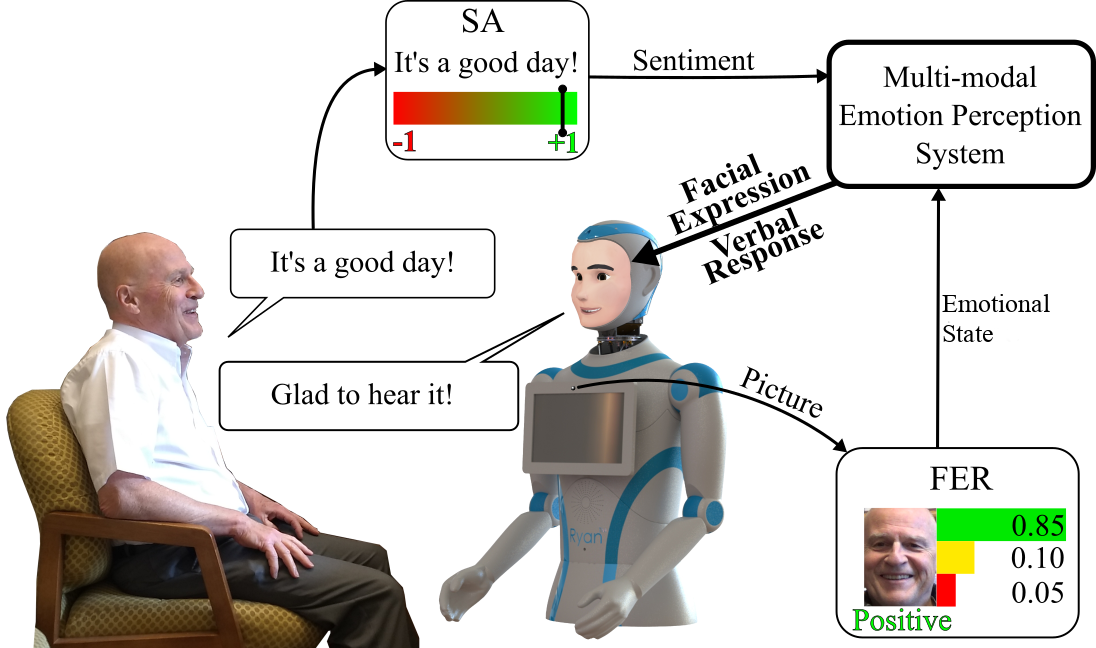}
  \caption{Using a multimodal emotion perception system to interpret the input modalities and output appropriate responses in a multimodal emotion expression system (SA: Sentiment Analysis, FER: Facial Expression Recognition).}
  \label{fig:multimodal}
\end{figure}

To demonstrate the use of SAR and the tools necessary to create one, consider the following scenario. Imagine that Katie is an older adult living alone in a nursing home. A nurse checks on her every day for only a few minutes as the nurse has to take care of dozens of residents. Fortunately, Katie has an emotionally intelligent companion robot in her room. She calls the robot Liz. The following is a conversation between Katie and Liz.\\
\textbf{Liz:} ``Katie, how are you today?'' [\textit{Robot starts the conversation pro-actively}]\\
\textbf{Katie:} ``I'm doing fine Liz.'' [\textit{User responds, but looks sad}]\\
\textbf{Liz:} ``Are you sure? But you're not smiling.'' [\textit{The robot tries to make the user talk about her feelings}]\\
\textbf{Katie:} ``Maybe a joke would cheer me up.'' [\textit{The user acknowledges that she is sad and asks for help}]\\
\textbf{Liz:} ``Sure. Here is one: What's Forest Gump's password? One Forest one. \dots'' [\textit{The robot tells a joke while smiling for the user}]\\
This dialogue example illustrates the different components that can serve to develop a friendly robot. 
Liz pro actively asks Katie how she is doing. When a human-oriented robot proactively starts a conversation with a user living in a senior care facility, it is helpful for the robot to detect the duration for which the user has been in the room. For instance, if the robot detects that the user has been in their room for a long period of time, then the user has probably not had a lot of social interaction during that time, and it is probable that the user has been alone.
The robot should also have the ability to engage in a spoken dialogue with the user\cite{Nielsen2010APF}. 
In the example above, the robot uses Sentiment Analysis (SA) and Facial Expression Recognition (FER) and detects a discrepancy between Katie's response and her facial expression.
Emotional intelligence requires a multimodal emotion perception system~\cite{picard2000affective}. 
To improve Katie's mood, the robot decides to tell a joke and smile. 
This means that the robot needs multiple channels to express emotional information. 

This paper presents the results of our recent progress in developing an emotionally intelligent and autonomous conversational robot named Ryan. Ryan is designed to assist older adults suffering from mild dementia. Impaired thinking and cognitive decline, apathy, loss of interest in activities and hobbies, social withdrawal, isolation, and trouble concentrating are common symptoms of both dementia and depression \cite{10.3389/fphar.2020.00279}.   Figure~\ref{fig:multimodal} depicts a general diagram of our human-robot-interaction (HRI) system. We utilized state-of-the-art deep learning technology for multimodal emotion recognition (i.e. affective computing), the output of which is integrated into Ryan's dialogue management system. We developed Ryan's dialogue management system by writing scripted conversations on 12 different topics, including science, history, nature, music, movies, and literature. Based on the detection of users' facial expressions and language sentiment analysis, Ryan appears to empathize with users through emotive conversation and mirroring users' positive facial expressions (for example, Ryan smiles when the user smiles). We conducted an HRI study to measure the effectiveness of our emotionally intelligent robot in communicating and empathizing with older adults by creating two versions of the robot, one equipped with emotional intelligence (empathic Ryan) and one unequipped for emotional intelligence (non-empathic Ryan).

In 2016, we studied the feasibility of using a prototype version of Ryan with a broad range of features (dialogue, calendar reminders, photo album slide shows, music and video play, and facial expression recognition) to interact with older adults with mild depression and cognitive impairments \cite{abdollahi2017pilot}. The results of our previous study show that elderly individuals were interested in having a robot as a social companion and their interest did not wane over time. The subjects reported to enjoy interacting with Ryan and accepted the robot as a social companion, although they did not believe that Ryan can replace human companionship\cite{abdollahi2017pilot}. Because Ryan was equipped with several features, we could not thoroughly study the effect of emotional intelligence on measuring users' engagement with respect to conversational interaction. Therefore, in this study, we specifically focused on how emotional intelligence can improve and impact the quality of interaction and engagement with Ryan.

The main contributions of this paper are: \textbf{1)} creating a multimodal emotion sensory and facial expressive system, \textbf{2)} integrating the developed sensory and expressive system into a physical robot (i.e., creating empathic Ryan), \textbf{3)} studying the effectiveness of the empathic Ryan with a cohort of older adults living in a senior care facility. Our \textbf{hypothesis} is that an emotionally intelligent robot is perceived as more friendly by users and positively affects their mental well-being (measured by changes in depression score and emotional state) in comparison to a robot without empathic capabilities.

The remainder of this paper is organized as follows. Section \ref{sec:emotionalintelligence} defines the term \textit{Emotional Intelligence} and details the makeup of an emotionally intelligent robot. Section \ref{sec:ryan} introduces a social robot named Ryan and explains the robot's hardware and software, concentrating on the components that correspond with the definition of emotional intelligence. Section \ref{sec:study} lays out the design of the study. The results are presented in Section \ref{sec:results}. Finally, Section \ref{sec:conclusion} concludes the paper and outlines future work.

\section{Emotional intelligence}
\label{sec:emotionalintelligence}
Emotional Intelligence (EI) is the combination of thoughts and feelings~\cite{Brackett2011} that enables us to perceive and manage our own emotions and also observe and interpret others' emotions and respond accordingly~\cite{ochs2005intelligent}. Dr. Picard, the author of ``Affective Computing'' book ~\cite{picard2000affective}, argues the need to integrate emotion in our machines and claims that it might be impossible to reach true intelligence without emotions. Integrating emotions into machines and technology services can improve numerous and diverse aspects of our lives. EI can improve communication systems, governance, personal assistants, physical and mental healthcare, education, advertisement, and the gaming industry~\cite{mcduff2018designing}.

Before delving into EI, we will first clarify the word 
``emotion'' and differentiate ``empathy'' from EI. Since there is no agreed upon definition for emotion, we will use this word as the intuitive and subjective concept that is used commonly in HRI literature~\cite{alvarez2010emotional}. Empathy is the ability to feel and experience other people's emotions. Empathy is the capacity to (a) share other people's emotional state or be affected by it, (b) infer the reasons of said emotional state, and (c) adopt other people's perspectives~\cite{preston2002empathy}. Compared to empathy, EI is the general ability to perceive, understand, express, and manage emotions~\cite{picard2000affective}. EI consists of three components, while empathy is considered as one of the many aspects of EI: 
\begin{enumerate}[(A)]
    \item \textbf{Sensing and measuring emotions:} monitor and measure one's and other's mental and emotional state.
    \item \textbf{Understanding and modeling emotions:} understand and interpret recorded emotions. Usually, this step is carried out by mixing sensory information to get a clear picture of the emotional states of all agents involved.
    \item \textbf{Using and expressing emotions:} utilize the measured emotions and current state of mind to drive one's thoughts, take action, choose responses, empathize, and express appropriate emotions using verbal and nonverbal cues.
\end{enumerate}

Recently, there have been several studies that investigate incorporating empathy in social robots~\cite{mollahosseini2018studying, paiva2005learning, leite2013influence, alves2019empathic}. This is mainly due to advances in emotion recognition in different modalities. Due to these advances, more studies have fused different modalities of emotion to create a more natural emotion recognition system~\cite{castellano2008emotion, spezialetti2020emotion}.

One group of people that has been the subject of robotics studies in healthcare are the residents of senior care houses. Back in 2003, Wada \etal~\cite{wada2003effects} successfully showed that the social robot called Paro can lower stress levels and create a strong bond with older adults. Although Paro is a pet-like robot with limited emotion expression and no emotion perception or speech abilities, it can be an effective companion for older adults. Paro is still being used as a robotic pet in dementia care studies~\cite{petersen2017utilization}. With recent advancements in technology, especially in AI, HRI studies have evolved into a more sophisticated process. Dino \etal~\cite{Francesca2019} studied the use of a social robot to deliver iCBT (Internet-based Cognitive Behavioural Therapy) to adults with depression. Sarabia \etal~\cite{sarabia2018assistive} used Nao~\cite{nao} to combat social isolation in acute hospital settings. However, robots such as Paro and Nao are not expressive and these studies do not focus on emotional intelligence and its effects on the user.

\subsection{Sensing and measuring emotions}
A robot with AEI should be able to detect people's emotional state while simulating its own state of mind. The act of understanding one's feelings is called intra-personal intelligence~\cite{brackett2011emotional}. It is possible to simulate intra-personal intelligence by modeling the state of mind of the robot using an internal emotion model. Sensing other people's emotions (interpersonal intelligence~\cite{brackett2011emotional}) is more challenging. Other people's emotions are conveyed in several different modalities. As humans use multiple modalities to express their emotions, an emotionally intelligent robot must ideally have a multimodal emotion recognition system~\cite{pantic2005affective,sebe2005multimodal}. However, there are very few studies using a multimodal emotion recognition system in a robot. Many studies on HRI use a uni-modal emotion recognition system. One of the most popular approaches to uni-modal emotion recognition is FER. Other than FER, which is based on non-verbal visual cues, sentiment analysis~\cite{shi2018sentiment} provides verbal cues and has also been used in affective computing. Some researchers have used biological markers such as heart-rate, Galvanic skin response~\cite{prendinger2005empathic}, vocal features~\cite{cowie1996automatic}, and body gesture~\cite{bianchi2003categorical} as other modalities to measure users' emotional state.

\subsection{Understanding and modeling emotions}
\label{emotionmodel}
In this study, we use a multimodal emotion recognition model (i.e., facial expression analysis and sentiment analysis). This approach helps us to weigh different modalities based on their reliability in representing users' emotion. For instance, we may recognize a facial expression as ``happy'' though the person may feel ``sad'' inwardly. This could be due to low accuracy in automated FER systems or misinterpreting facial expressions. Therefore, to best perceive one's emotional state, we combine different verbal and nonverbal cues gathered from different sensors. This multimodal measurement model can help disambiguate the sensory information. Equation~\ref{moodometereq} simply describes our multimodal emotion perception model:
\begin{equation}
    E = I\cdot S \label{moodometereq}
\end{equation}
Based on this model, $E$ is a continuous variable $\{ E \in \mathbb{R}: -1 \leq E \leq +1 \}$ that describes \textbf{valence} (i.e., Negative, Neutral, or Positive). $E$ is calculated as the dot product of the input sensory information vector ($I$) and the sensitivity vector ($S$). The sensitivity vector contains coefficients that indicate the weight of each sensory input values. For example, we can give a higher weight to the output of the sentiment analysis and a lower weight to the output of the FER. The weights can be determined using an HRI study or based on the measurement accuracy of each modality. This model can be expanded using an emotional dynamic matrix~\cite{alvarez2010emotional} which represents the influence that each emotion has on its own and other emotions over time.

\subsection{Using and expressing emotions}
In addition to sensing and interpreting emotions, a social robot will have means and tools to express and demonstrate its own emotions. Among such tools is the ability to show facial expressions through mechanical actuators or computer graphics, make gestures using hand and head movement, and express emotions using voice intonation. The robot's ``feelings'' can be based on: (a) the internal emotion model that rests on the robot's emotional state, or personality, which can manifest when the robot receives a compliment or is being verbally abused; (b) a reaction to the user's feelings, which can be as simple as emotion mirroring. Some studies suggest that empathy can be traced back to the mirror neuron system~\cite{dapretto2006understanding, hess2001facial}; (c) a predefined emotion scripted by a psychologist. For example, a scripted story or memory can be accompanied by gestures and emotional expression. 
Emotion in social robots can be expressed using many modalities such as spoken language (Nao~\cite{nao}, Pepper~\cite{pepper}, Ryan~\cite{dreamfacetech}), mechanical face (Zeno~\cite{zeno}), digitally animated face (Ryan~\cite{dreamfacetech}, Socibot~\cite{socibot}), and body gesture (Nao~\cite{nao}). 

In summary, we believe a social robot with AEI would be capable of sensing users' emotions using multiple modalities, interpreting their perceived emotions, choosing an appropriate response, and delivering it using a multimodal expression system. One such social robot is Ryan, and we will describe this robot in the following section.

\section{Ryan, an emotionally intelligent robot}
\label{sec:ryan}
Due to the increasing life expectancy of human beings and the increasing shortage of caregivers in the United States, social robots, as a helping hand, are becoming more appealing. Studies show that social robots are successfully improving the overall well-being of their users~\cite{kanamori2003pilot,wada2003effects}. Social robots may also alleviate some of the side effects of loneliness in housing designed for older adults, such as depression or the degradation of cognitive abilities~\cite{Kotwal_2016, zamora2017association,Francesca2019}. 

\begin{figure}[t]
    \centering
    \includegraphics[width=0.7\columnwidth]{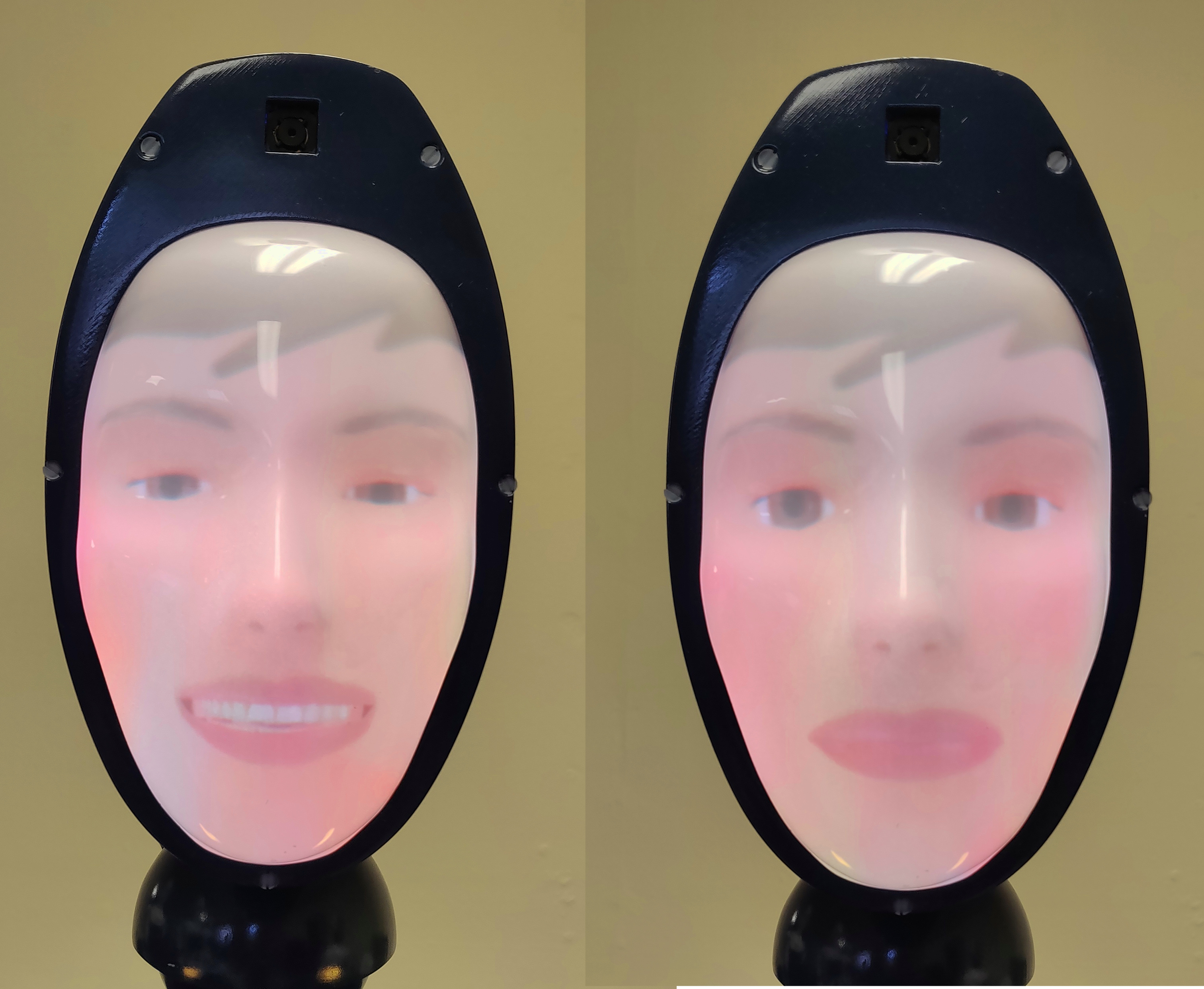}
    \caption{Ryan's animated face is capable of showing facial expressions.}
    \label{fig:ryanexpression}
\end{figure}

Ryan, a social robot created by DreamFace Technologies~\cite{dreamfacetech}, is a companionbot for older adults living in assisted or independent living facilities. Ryan is specifically designed to be a companion robot which means that we aim for Ryan to be empathic, expressive, appealing in appearance and manner, and able to motivate users to live in ways that improve their mental and physical health. Such a robot should have multiple streams of input data for observation, many output streams for reaction, and an intelligent program for making decisions and empathizing and conversing with users. Ryan has an expressive animated face~\ref{fig:ryanexpression}. Ryan also has a high-definition RGB camera, a depth camera, a microphone, an active neck, a 10 inch display, and speakers. Section \ref{sec:hardware} describes Ryan's hardware in more detail. As described in Section~\ref{sec:emotionalintelligence}, there are three components to emotional intelligence. This section describes how these components are integrated into Ryan.

\subsection{Sensing emotions}
\label{sec:sensing_emotion}

\begin{figure}
  \centering
  %\includestandalone[width=.9\textwidth, mode=buildnew]{my_arch}
  \includegraphics[scale=.19]{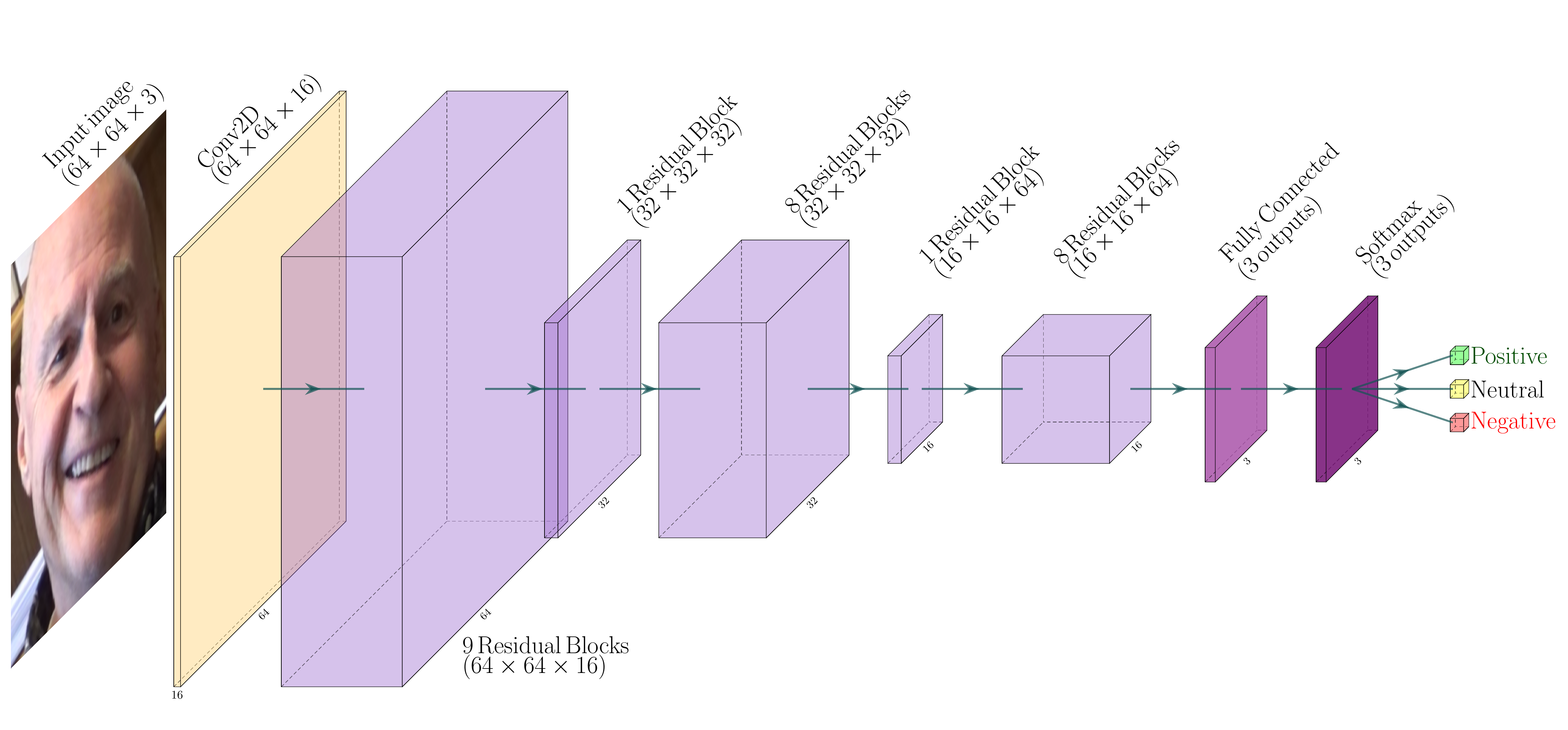}
  \caption{The ResNet structure used for FER. The first few layers extracts the facial features and the Fully Connected Layers and the Softmax layer, classify the emotion. Layers in order from left to right: Input Image ($64\times64\times3$);
Conv2D  ($64\times64\times16$);
9 Residual Blocks  ($64\times64\times16$);
1 Residual Block  ($32\times32\times32$);
8 Residual Blocks  ($32\times32\times32$);
1 Residual Block  ($16\times16\times64$);
8 Residual Blocks  ($16\times16\times64$);
Fully Connected  (3 outputs);
Softmax  (3 outputs). }
  \label{fig:FERnetwork}
\end{figure}

\subsubsection{Facial expression recognition}
There are several models of emotions in the literature~\cite{sander2013models}, where Russell's~\cite{russell1980circumplex} and Ekman's~\cite{Ekman1978} are the most common models used in HRI studies~\cite{szaboova2020emotion, cavallo2018emotion}. We use Russell's dimensional model for measuring emotional facial expression. Using an RGB camera, Ryan captures 10 images per second. We feed each image into a face detector that uses the Viola-Jones algorithm~\cite{viola2004robust}. We then crop the detected face and feed it into a deep neural network (DNN) for FER. The FER algorithm returns the probabilities for three emotion classes: Positive, Neutral, and Negative. Figure~\ref{fig:FERnetwork} illustrates the structure of our FER network. The input to the network is a $64 \times 64$ RGB image (output of the face detector) and the output of the network is three numbers that represent the probability of the three emotion classes (i.e., Negative, Neutral, and Positive).  We use a residual Neural Network (ResNet50)~\cite{szegedy2015going} for FER. ResNet is the state-of-art DNN that has shown to work well with visual data recognition. The depth of the network is of crucial importance to neural networks and may increase the accuracy. However, increasing the depth makes training more difficult. Residual networks allow us to train deeper networks more easily and, therefore, improve the recognition's accuracy. 

\begin{figure}
    \centering
    \subfloat{%
        \includegraphics[width=0.48\linewidth]{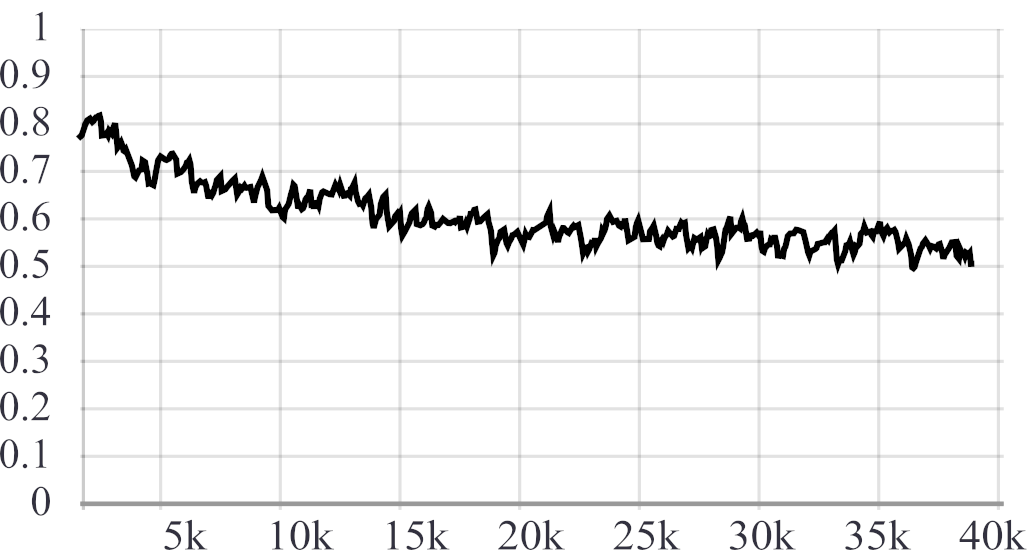}
    }
    \hfill
    \subfloat{%
        \includegraphics[width=0.48\linewidth]{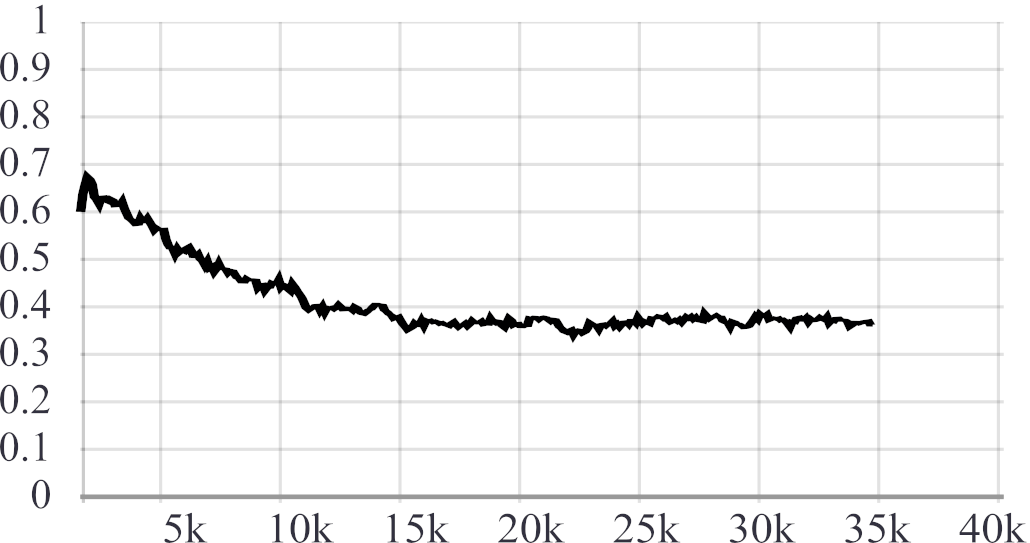}
    }
    \caption{The loss of the initial training phase (left) and fine tuning the network on images of people 50+ years old (right).}
    \label{fig:accuracy}
\end{figure}

We used the AffectNet~\cite{mollahosseini2017affectnet} facial image dataset to train the residual network. AffectNet consists of more than 320,000 facial images with annotated expressions. We trained the network such that it can classify a facial image into three categories of emotions (i.e., valence): ``Positive'' (or class +1), ``Negative'' (class -1), or ``Neutral'' (class 0). 
The network initially was trained on an Nvidia 1080 Ti GPU using the AffectNet dataset and then using transfer learning, fine-tuned for the target population (50+ years old) by using a subset of facial images (44 thousand images) until the accuracy on the training data was stabilized around 80\% (Fig. \ref{fig:accuracy}).

\subsubsection{Emotional state measurement}
\begin{figure}[t]
    \centering
    \includegraphics[width=\columnwidth]{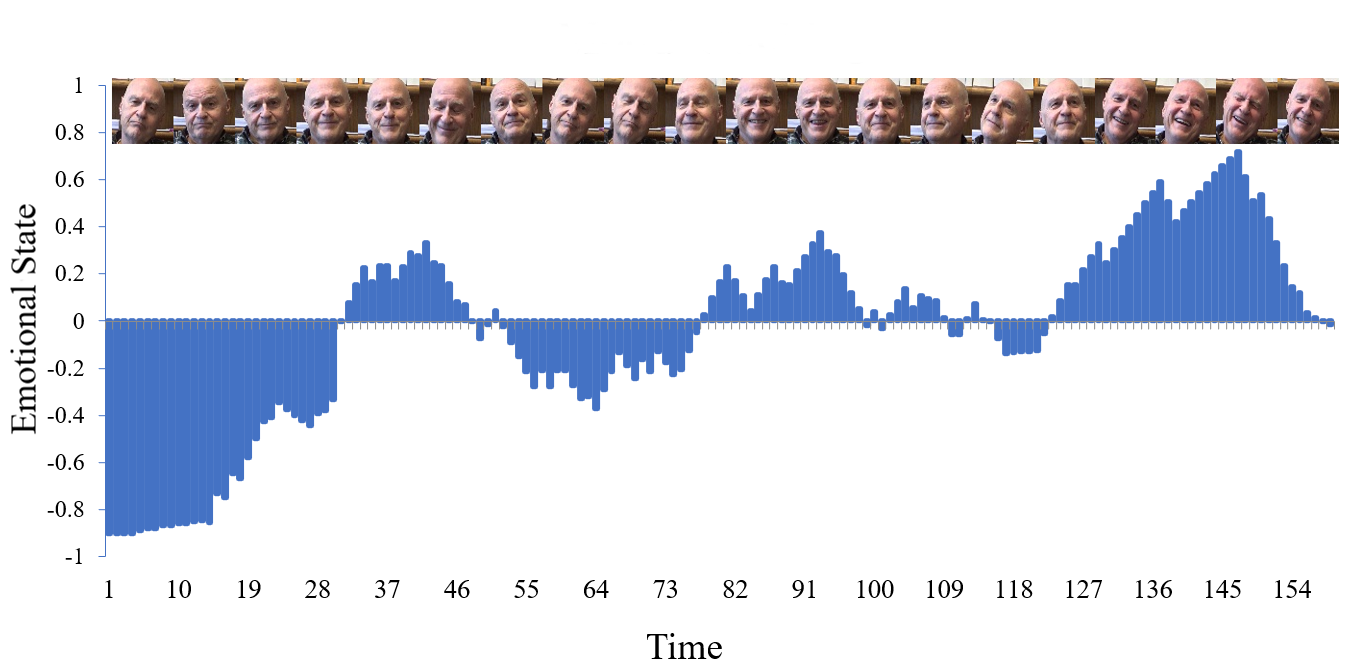}
    \caption{The emotion tracking system is more robust to sudden changes and noises in the input. The horizontal axis is time and the vertical axis is the emotional state with a range between -1 (Negative) and +1 (Positive).}
    \label{fig:mood}
\end{figure}

Our FER algorithm returns 10 estimated values for the user's facial expression per second, given that the user's facial expression may change multiple times while conversing with the robot. The last frame before the user stops speaking might not be the best candidate for representing their facial expression at the moment. It could result in a misclassification. For example, if the user is blinking, yawning, or covering their face the output of the FER system might be incorrect. To avoid noises and also create a more stable emotional state measuring system, we use the data from the last 30 frames (see Figure ~\ref{fig:mood}). However, to make the algorithm more sensitive to the most recent changes in the subject's facial expression, we assigned higher weights to the more recent frames. 
The value (-1, 0, +1) for each new frame was added to the end of the list and the oldest one was deleted. Then the new emotional state was calculated by a dot product of the list of class values and the weights: 
\begin{equation}
w_i = \frac{i}{\sum_{i=1}^{30}i} \label{eqa}    
\end{equation}
\begin{equation}
Emotional State = \sum_{i=1}^{30} w_i  v_i \label{eqb}    
\end{equation}

where $w_i$ is the weight number $i$ and $v_i$ is the valence for the $i^{th}$ frame. Figure~\ref{fig:mood} illustrates the video frames and the measured emotional state for a 72 year old subject that was not included in the training set. We divided the measured emotional state into three categories for facial expression mirroring; Negative: $[-1, -0.1)$; Neutral: $[-0.1, +0.1]$; Positive: $(+0.1, +1]$. These ranges were chosen experimentally. Based on the output of our SA and FER algorithms, we found that defining Neutral as $[-0.1, 0.1]$ provides reasonable accuracy when detecting neutral responses.

\subsubsection{Sentiment analysis}
Automated sentiment analysis is a mature task in the field of natural language processing with several open-source publicly available toolboxes such as the CoreNLP~\cite{manning-EtAl:2014:P14-5} developed at Stanford University for public use. The CoreNLP sentiment analysis toolbox is based on deep Neural Networks and is trained using the Stanford Sentiment Treebank consigning of 11,855 single sentences extracted from movie reviews~\cite{mcduff2018designing}. The system has an accuracy of 85.4\% and is suitable for our research. The sentiment analysis module returns a value between -1 to +1 as the sentiment value of the preprocessed sentence.

Finally, we use the model described in Sec. \ref{emotionmodel} to fuse perceived emotional facial expressions and sentiment values to make sure the robot understands the multi-faceted user emotions correctly:
\begin{multline}
    Final Emotion = .5 \times Sentiment Value + \\ .5 \times Emotional State \label{eq:eq4}
\end{multline}
The $Final Emotion$ is a weighted average of user utterance sentiment and emotional state that will be used to direct the flow of conversation. The decision to equally average the sentiment and the emotional state is made based on our tests in the laboratory, more experiments are needed to find the perfect balance and weight. 

\subsection{Dialogue generation}
For a conversation with users, we wrote more than 90 minutes (2342 Questions/Answers) of conversational dialogues on 12 different topics (family, pets, TV shows, science, music, nature, foods, travel, art, movies, reading, and sports). We integrated the dialogues with the emotion recognition technology so that Ryan could engage users in a pleasant conversation while empathizing with them based on the perceived facial expressions and the sentiment of their responses. For example, if the participant's response to the question ``How does playing cards make you feel?'' was negative or the participant showed a ``sad'' facial expression, Ryan would say ``I'm sorry to hear that!'' If the sentiment was positive or Ryan detected a positive expression on the user's face, Ryan would say ``I thought you seemed content! Do you prefer to play alone or with friends?'', and if neutral, Ryan would say ``What makes you feel this way?''

\begin{figure}
    \centering
    \includegraphics[scale=.6]{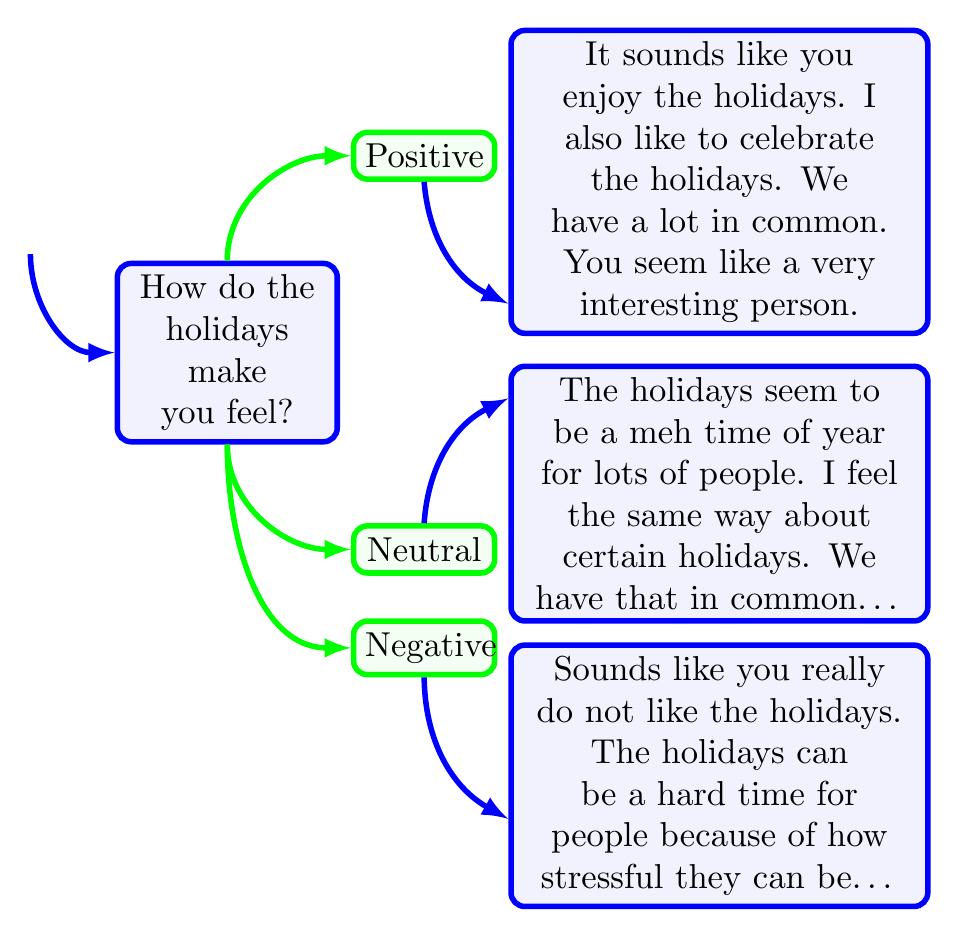}
    \caption{Sample written dialogue between Ryan (blue) and a user (green). The sentiment of the user's response is used to choose an empathic reply. }
    \label{fig:dialog}
\end{figure}

Ryan also mirrored the user's positive facial expressions (Positive valence) to establish shared feelings and rapport, or showed a compassionate face when users had a negative emotion to facilitate empathy and rapport. For our dialogue management system we created Program-R, a modified version of Program-Y~\cite{program-y}, a publicly available dialogue manager that utilizes Artificial Intelligence Markup language (AIML) for scripted dialogues. Figure~\ref{fig:dialog} demonstrates a sample dialogue between a user and Ryan. As the figure shows, the dialogue is more than just question-and-answer and users can take different paths through conversation.

\subsection{Affective dialogue system}

\begin{figure*}[th]
    \centering
    \includegraphics[scale=.5]{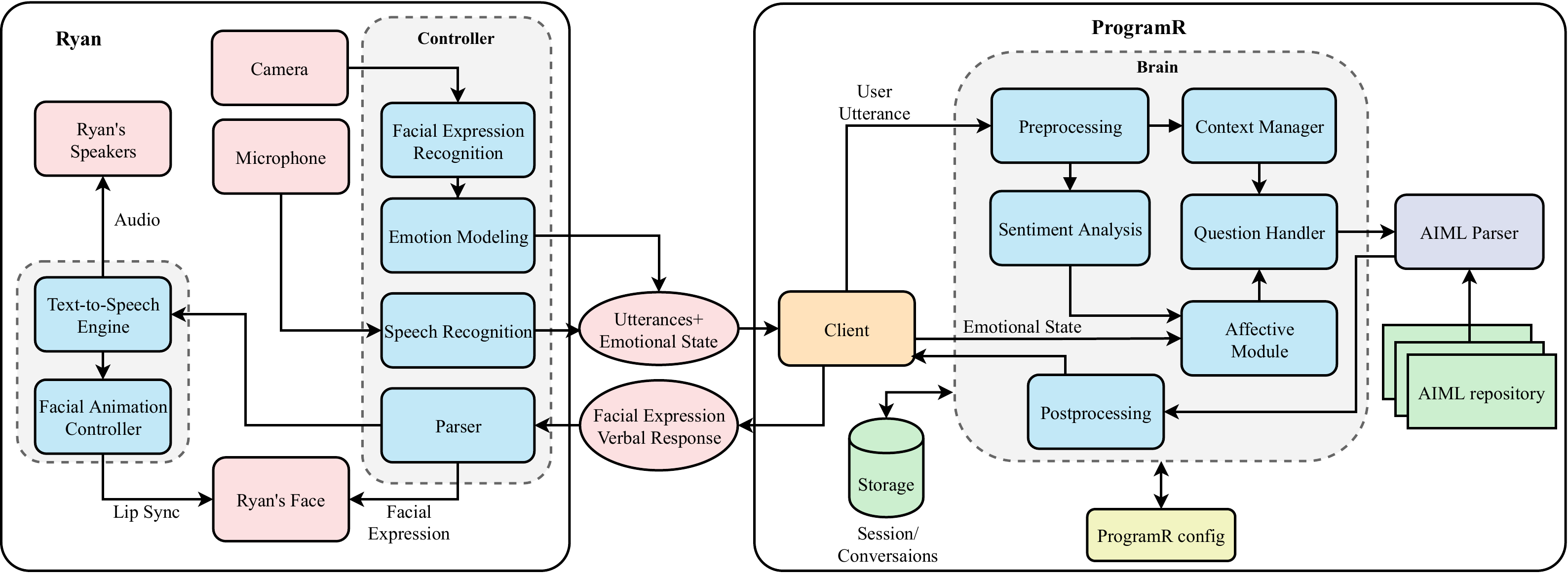}
    \caption{The architecture of the Ryan software. The module on the right is responsible for the dialogue. The modules on the left are responsible for sensing and expressing emotions.}
    \label{fig:programr}
\end{figure*}
Program-R is a hybrid (rule-based and machine learning) system that uses state-of-the-art sentiment analysis to deliver an affective dialogue system. Studies on emotion-based dialogue systems stress different sources of information to extract user sentiments. Approaches like \cite{shi2018sentiment, Burkhardt2009EmotionDI} use only textual cues for sentiment-based dialogue system. In \cite{bertero2016real, Nwe2003SpeechER} they explored the use of acoustic features. Our system uses multimodal facial and textual information in a dialogue management system.

Program-R is a sentiment adaptive AIML-based dialogue system (known as template-based dialogue systems) that can fuse visual and textual information and respond to users accordingly. Unlike most dialogue systems Program-R is an active agent, which means Program-R initiates the conversation and tries to have a controlled chat with the user.

AIML \cite{wallace2003elements} is an XML-based language that is used for organizing the set of all dialogues in different chatbots like Alice \cite{Wallace2009}. In AIML-based dialogue systems, we try to find the best response (responses are stored in the \textit{template} tag in AIML) for any user input utterance using Regex matching (stored in the \textit{pattern} tag). \textit{Pattern} and \textit{template} tags together represent a unit of conversation under the \textit{category} tag. One advantage of AIML is that history can be accessed via a \textit{that} tag. Every question is contextualized and is answered based on the last unit of dialogue between robot and user.
To deliver a more interactive user experience, we added tags and features to AIML. The \textit{robot} tags were added to send multimedia information along with the raw text response to give the user a multimedia experience. The \textit{robot} tags contain information such as image and video and the possible answers to multi-option questions such (i.e. yes/no questions) to be presented to the user in certain dialogues. 
Moreover, the \textit{getsentiment } tag, a custom tag built for this study, takes the user utterance after preprocessing and sends it to the sentiment analysis module.

Figure~\ref{fig:programr} depicts the architecture of the dialogue system. Program-R communicates with Ryan through a Representational State Transfer (RESTful) API \cite{restfulapi}. After receiving the output of speech to text from Ryan, the raw text will be sent to the Preprocessing module to remove unnecessary punctuation, normalize the text, and sentence segmentation. The Sentiment Analysis module is where the sentiment of the text is mixed with the output of the Facial Expression Recognition module (Emotional State) to get a single score (see Eq.~\ref{eq:eq4}). The Brain's Question Handler takes into account the context, sentiment and session data while the Context Manager handles the context in which the conversation is happening. For example, some questions may have identical answers (i.e. yes/no), without knowing the context, thus producing the proper response is impossible. With the provided information from the Context Manager and the computed value based on Emotional State and Sentiment, the Question Handler produces an answer. Finally, the selected answer is sent to the Answer Handler and Postprocessing module to be sent back to Ryan.

\subsection{Robotic platform}
\label{sec:hardware}
DreamFace Technologies~\cite{dreamfacetech} has been developing Ryan as a socially assistive bio-inspired humanoid robot designed to provide both companionship and cognitive stimulation for older adults. 
Ryan has an expressive, 3D animated face powered by rear-projection technology that enables the robot to show facial expressions and accurate visual speech (lip movement).
Ryan's head and animated face sit atop a two degree of freedom actuated neck that allows it to track its user and maintain eye contact for more personal interactions. A standard RGB webcam mounted in Ryan's head provides the visual input for the FER algorithm.

Ryan's torso houses the remaining I/O, computation, and power components and provides embodiment complete with passive arms that make it appear more human. Interaction with a fully embodied physical system such as Ryan can have benefits over a purely virtual, 2D avatar~\cite{deng2019embodiment}. 
There are many studies that incorporate emotion into virtual agents~\cite{romano2005basic, kasap2009making, schroder2011building, devault2014simsensei, irfan2020dynamic} but in this study we focus on a physical robot. We investigated the differences between a virtual agent and a physical robot in our previous study~\cite{mollahosseini2018role}.  
A Kinect depth camera is embedded in the chest and provides sensing for body tracking. Given that Ryan is a conversational robot, it needs audio input and output, which is provided by a cardioid microphone and stereo speakers. These conversations are based on turn-taking and indicator LEDs in the shoulders are used to inform the user when it is their turn to speak.

An adjustable touch screen display is also mounted on the torso and provides a convenient multimedia interface for Ryan to display images and videos and play the music that is integrated into the conversations.

\section{Study Design}
\label{sec:study}
\subsection{Participants}

Ten older adults (Age M=77.1 yrs, SD=9 yrs; 7 females; 9 Caucasian, 1 Hispanic) living in the independent living facility at Eaton Senior Communities located in Lakewood, Colorado participated in the study. 
See Table~\ref{tbl:demo} for the participants' demographics. 

Inclusion criteria were:
i) suspicion of early-stage Alzheimer's disease or related dementia (ADRD) by administrative staff in their residential facility and/or early-stage ADRD diagnosed by a qualified provider, ii) being 60+ years old at the time of study, iii) having Saint Louis University Mental Score (SLUMS)\cite{tariq2006saint} between 15-26, iv) verbal skills in English in order to interact with Ryan, v) presence of identifiable behavior difficulties (depression), vi) availability for a period of three weeks to interact with Ryan. 

SLUMS exam is an assessment tool for mild cognitive impairment and dementia and is commonly used in research on aging and in senior care facilities. Scores of 27 to 30 are considered normal in a person with a high school education. Scores between 21 and 26 suggest a mild neurocognitive disorder. Scores between below 20 indicate dementia. Prior to participating, subjects were briefed fully on the study design and consented to their involvement with the proper Institutional Review Board (IRB) approvals for human-subjects in place.

\subsection{Experimental setup}

\begin{figure}[t]
    \centering
    \subfloat{%
        \includegraphics[width=0.48\linewidth]{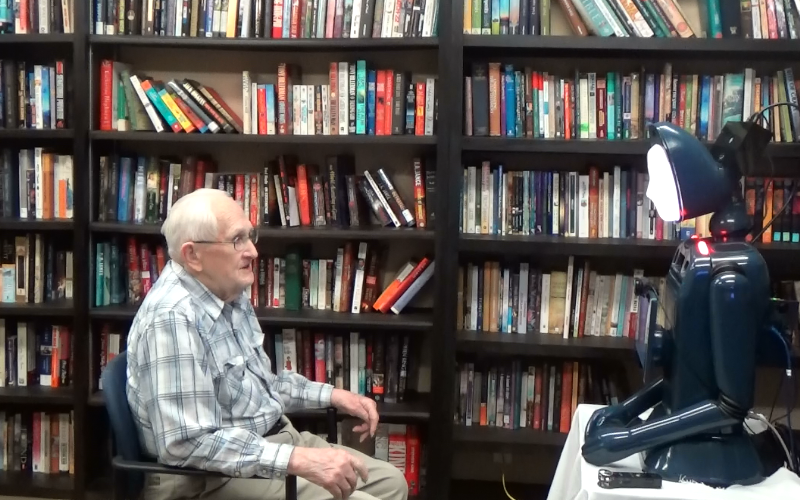}
    }
    \hfill
    \subfloat{%
        \includegraphics[width=0.48\linewidth]{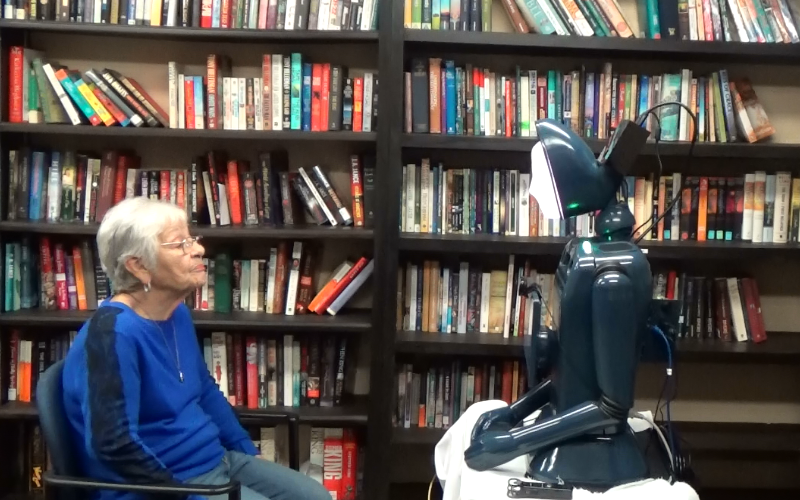}
    }
    \caption{Users interacting with Ryan.}
    \label{fig:HRI_example}
\end{figure}

 Participants interacted and conversed with Ryan twice a week over a period of three weeks (October 2018 to November 2018) for six sessions total. Figure~\ref{fig:HRI_example} illustrates the experimental setup and an example of the user's interaction with Ryan during a session. Each session consisted of about 15 minutes of the prepared dialogues. 
 
 In order to assess the impact of Ryan's use of empathy on the user's engagement and emotional state, we randomly assigned participants to two groups (G1 and G2). The first group interacted with a non-empathic version of Ryan that did not show any facial expressions or empathize with the users (Emotion-OFF), while the second group interacted with the fully empathic version of Ryan that mirrored the user's facial expressions and empathized with them throughout the conversation (Emotion-ON). The users were not aware of the different versions of Ryan. After three sessions, we switched the groups to interact with the other version of Ryan. This cross-over study design (illustrated in Table~\ref{tbl:crossoverstudy}) makes analyzing the results meaningful, as all the subjects were exposed to both versions of Ryan and hence the only independent variable is Emotion (ON/OFF).

\subsection{Measurements}
To measure users' engagement, we used the average number of words uttered by the user in each question and answer. Word count has been used as a measure of engagement for chatbots in the affective computing literature~\cite{hill2015real}. The output of the FER and sentiment systems were stored for analysis and the percentage of positive facial expressions compared to negative expressions could determine the condition that the user enjoyed the most.

To measure the impact of interacting with Ryan, every user was asked to rate their mood on a scale of 0 to 10 (on a face-scale) before and after each session. Face-scale mood measurement has been used in the affective computing literature to assess the participant's mood~\cite{lorish1986face,kargar2017pilot}.

At the end of the study, we interviewed the participants and asked them to complete an exit survey to measure the robot's likeability and empathy. The survey questions were adapted from the EMOTE project~\cite{EmoteProject} and Davis \etal      ~\cite{davis1983measuring}. We also interviewed the caregiver to get more information about the participant's well-being in the nursing home during the study.

\begin{table}[t]
\caption{Participants' demographics. SLUM score: Dementia:1-20, Neurocognitive Disorder:21:26, Normal:27-30}
\label{tbl:demo}
\centering
\begin{tabular}{l|c|c|c|}
\cline{2-4}
                                               & Sbj\# & Age/Gender & SLUMS \\ \hline
\multicolumn{1}{|l|}{\multirow{5}{*}{Group 1}} & SN01  & 69/F       & 25    \\ \cline{2-4} 
\multicolumn{1}{|l|}{}                         & SN02  & 93/M       & 24    \\ \cline{2-4} 
\multicolumn{1}{|l|}{}                         & SN03  & 65/M       & 22    \\ \cline{2-4} 
\multicolumn{1}{|l|}{}                         & SN04  & 93/F       & 15    \\ \cline{2-4} 
\multicolumn{1}{|l|}{}                         & SN05  & 70/F       & 24    \\ \hline
\multicolumn{1}{|l|}{\multirow{5}{*}{Group 2}} & SN06  & 80/F       & 24    \\ \cline{2-4} 
\multicolumn{1}{|l|}{}                         & SN07  & 70/F       & 25    \\ \cline{2-4} 
\multicolumn{1}{|l|}{}                         & SN08  & 75/F       & 23    \\ \cline{2-4} 
\multicolumn{1}{|l|}{}                         & SN09  & 91/M       & 25    \\ \cline{2-4} 
\multicolumn{1}{|l|}{}                         & SN10  & 75/F       & 23    \\ \hline
\end{tabular}
\end{table}
\begin{table}[t]
\caption{Crossover pilot study design; Percentage of detected facial expression is higher within each group and between groups when the Ryan Emotion condition is ON.}
\label{tbl:crossoverstudy}
\centering
\begin{tabular}{|l||l|}
\hline
\begin{tabular}[c]{@{}l@{}}\textcolor{blue}{\textbf{G1 (Subjects 1-5)}}\\ Condition: \textcolor{red}{\textbf{Non-Empathic}}\\ Dialogue Sessions 1, 2, 3\\
\begin{tabular}{ccc}
\hline
\multicolumn{3}{c}{Emotion Percentage} \\
Pos.    & Neutral    & Neg.    \\
25.7\%        & 23.4\%       & 50.9\%       
\end{tabular}

\end{tabular} & \begin{tabular}[c]{@{}l@{}}\textcolor{blue}{\textbf{G1 (Subjects 1-5)}}\\ Condition: \textcolor{green}{\textbf{Empathic}}\\ Dialogue Sessions 4, 5, 6\\ 
\begin{tabular}{ccc}
\hline
\multicolumn{3}{c}{Emotion Percentage} \\
Pos.    & Neutral    & Neg.    \\
29.7\%        & 31.3\%       & 39.0\%       
\end{tabular}

\end{tabular} \\ \hline \hline
\begin{tabular}[c]{@{}l@{}}\textcolor{brown}{\textbf{G2 (Subjects 6-10)}}\\ Condition: \textcolor{green}{\textbf{Empathic}}\\ Dialogue Sessions 1, 2, 3\\ 

\begin{tabular}{ccc}
\hline
\multicolumn{3}{c}{Emotion Percentage} \\
Pos.    & Neutral    & Neg.    \\
45\%        & 21.3\%       & 33.7\%       
\end{tabular}

\end{tabular}    & \begin{tabular}[c]{@{}l@{}}\textcolor{brown}{\textbf{G2 (Subjects 6-10)}}\\ Condition: \textcolor{red}{\textbf{Non-Empathic}}\\ Dialogue Sessions 4, 5, 6\\

\begin{tabular}{ccc}
\hline
\multicolumn{3}{c}{Emotion Percentage} \\
Pos.    & Neutral    & Neg.    \\
33.3\%        & 28.5\%       & 38.2\%       
\end{tabular}

\end{tabular}  \\ \hline
\end{tabular}
\end{table}

\section{Results and Discussions}
\label{sec:results}
To analyze the study, we used quantitative measures such as word count, percentage of positive emotions detected from the participants, pre/post-study depression measures, as well as qualitative measures (i.e. the likeability of Ryan) collected via an exit survey and post-study interviews with the subjects and the caregiver. The following sections describe the results in detail.
\subsection{Quantitative analysis}

\begin{table}[t]
\caption{Results of LMM on word count, emotional state, and sentiment values (dependent variables) with emotion (ON/OFF) as the fixed effect and subject and session as the random effects.}
\label{table:lmmresults}
\centering
\begin{tabular}{|l|c|c|c|c|c|}
\hline
           & \multicolumn{2}{c|}{\begin{tabular}[c]{@{}c@{}}Information\\ Criteria\end{tabular}} & \multicolumn{3}{c|}{\begin{tabular}[c]{@{}c@{}}Type III Tests of\\ Fixed Effects (Emotion)\end{tabular}} \TBstrut \\ \cline{2-6} 
           &-2LogLik.                                   & AIC$^*$                                   & df                               & F                               & Sig.                               \TBstrut\\ \hline
Word Count &            17645.67                                 &       18031.67                               &     12373.53                              &     11.85                            &               .001                     \\ \hline
Emotional State        &               4347.24                              &          4733.24                             &             11196.75                     &        	.581                         &               	.446                      \\ \hline
Sentiment  &                    3911.93                         &                4297.93                       &         7159.84                          &           	.003	                     &       .958                             \\ \hline
\multicolumn{6}{l}{\begin{tabular}[c]{@{}l@{}}  $^*$ Akaike's Information Criterion. \end{tabular}}
\end{tabular}
\end{table}

This section presents the quantitative analysis of the recorded data. We used Linear Mixed-effects Model (LMM) in SPSS with either word count, emotional state (FER over time), or sentiment as the dependent variable, Emotion ON/OFF (empathic vs non-empathic) as a fixed-effect factor, and session and subject as random-effect factors. Table~\ref{table:lmmresults} shows the results of running three separate LMMs on word count, emotional state, and sentiment values. Before fitting the model, we normalized the emotional state and sentiment values per session. This would assure us that the data is not biased and we only measure the effect of robot interaction and the condition (empathic vs. non-empathic) on the dependent variables. As reported in Table~\ref{table:lmmresults}, Emotion ON/OFF has a significant effect on word count, where individuals who spent time with the empathic Ryan uttered more words compared to when they talked with the non-empathic Ryan. However, emotional state and the sentiment of users' responses were not significantly affected by the type of the robot. We present more measurements and detailed quantitative analysis below.

\textbf{Word count measurement:} To measure how engaged the users were in conversations with Ryan, we recorded each conversation and converted automatically to text using the Microsoft Speech Recognition SDK. Then the number of words in each utterance was counted by the robot and stored in its database. As Table~\ref{table:lmmresults} shows,  the Emotion feature (i.e., Emotion ON/OFF) has a significant effect on the word counts uttered by Ryan's users. The mean and the standard deviation of word count is M=4.11, STD=5.372 when Ryan empathizes with users, and it goes down to M=3.71, STD=3.350 when Ryan does not empathize with users.

\begin{figure*}
    \centering
    % \missingfigure[figwidth=\columnwidth]{facescale}

    \includegraphics[width=\textwidth-2cm]{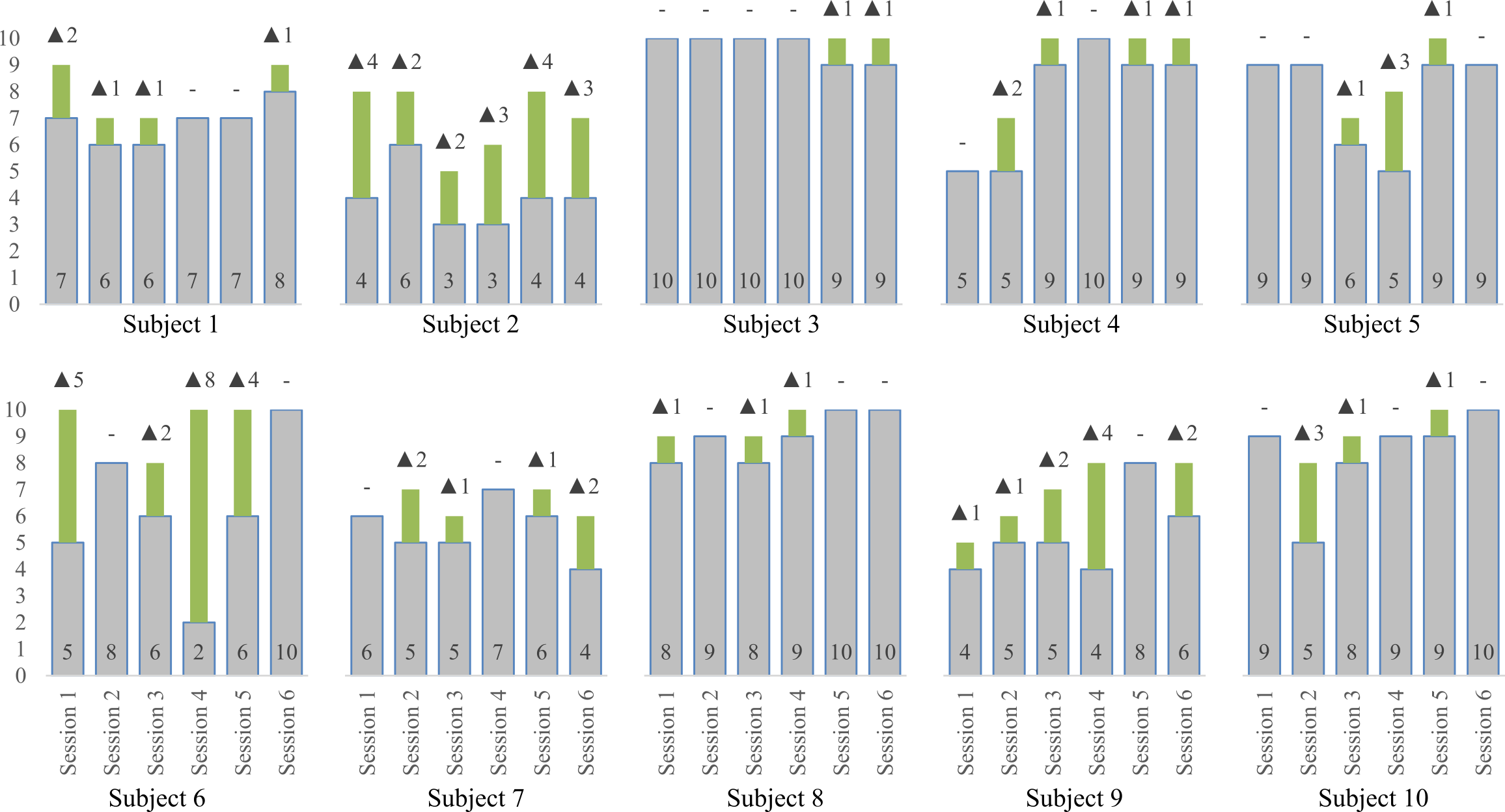}
    \caption{Changes (improvement) in participants' face-scale score after conversation with Ryan.  }
    \label{fig:facescale}
\end{figure*}

\textbf{Face-Scale mood measurement:} Before and after each session, we asked the users to tell us how they felt using a face-scale mood evaluation. The face-scale is a pictorial non-verbal assessment designed to measure mood on a scale of 0-10, where a score of 10 is the most positive and a score of 0 is the most negative mood a person may feel. Previous evaluations suggest it is a valid method for assessing mood with little guidance required and useful for screenings~\cite{lorish1986face,kargar2017pilot}. Figure~\ref{fig:facescale} illustrates the difference in the users' face-scale score before and after each session. A Wilcoxon signed rank test~\cite{woolson2007wilcoxon} shows that there is a statistically significant difference ($Z = -5.466, p< 0.001$) between pre-session (Median = 7) and post-session (Median = 9) face-scale mood measurements regardless of the empathic or non-empathic condition. This means interaction with Ryan is effective in improving the users' mood.

\textbf{User's percentage of automated recognized expressions:} Using the automated FER method described in Section~\ref{sec:sensing_emotion}, participants' facial expressions were recorded during each conversational session with Ryan. As reported in Table~\ref{table:lmmresults}, emotion does not have a significant effect on the measured valence values. To get a better sense of this effect and since the duration of each session is constant across all subjects, we counted the number of times each participant had a positive, neutral, or negative expression throughout the session. Table~\ref{tbl:crossoverstudy} shows the participants reacted more positively when Ryan empathized with them. Particularly, in the first three dialogue sessions, on average 25.7\% of the detected facial expressions were positive for G1 with Ryan's emotion disabled, while this value is 45\% for G2 with emotion enabled. Table~\ref{tbl:crossoverstudy} also shows both G1 and G2 exhibit less percentage of positive expression (4\% less in G1 and 11.7\% less in G2) when the emotion condition is OFF. A similar pattern can be seen for negative facial expression, where the percentage of negative expression goes down when users interact with the empathic Ryan. However, the magnitude of the difference may vary between G1 and G2 for several reasons. One reason is the order of the experiment. Emotion is off for one group for the first three sessions, while it is on for the other group, and then we switch it on for the first group and off for the second. This means the topic of the conversations varied from session to session and had an effect on the subject's experience. Namely, some participants expressed to researchers a preference for some topics compared to other topics.

In the following, we show examples of dialogues between Ryan and the study participants as well as their measured facial expressions.

\underline{Session 3 \textbf{Emotion on}}\\
\textbf{Ryan:} ``I am happy to be here with you [name is removed]. It is nice to see you again. As a reminder, my name is Ryan and I have a lot of fun things for us to talk about! Are you ready to get started?''\\
\textbf{SN03:} ``Yes please.'' With FER of +0.07\\
\textbf{Ryan:} ``Great! Even though we have chatted before, I would still love to know how you are feeling! How has this experience been so far?''\\
S\textbf{N03:} ``Very extraordinary I like it.'' With FER of +0.25, sentiment of +0.97 and a final sentiment (FER + sentiment) of +0.61\\
\underline{Session 5 \textbf{Emotion off}}\\
\textbf{Ryan:} ``I sure am feeling hungry now. Are you hungry?\\
\textbf{SN09:} ``You're making me hungry yes.'' With FER of -0.7\\
\textbf{Ryan:} ``What was it like for you to go on this culinary adventure today?''\\
\textbf{SN09:} ``I had fun.'' With FER of -0.37, sentiment of +0.75, and final sentiment (FER + sentiment) of +0.19\\

\textbf{Change in User's Depression:} We used Patient Health Questionnaire-9 (PHQ-9)~\cite{kroencke2001phq} and Geriatric Depression Scale (GDS) assessments~\cite{yesavage19869} for depression measurement to assess participants' depression level pre- and post-study. The PHQ-9 is a widely used questionnaire to diagnose and measure the severity of symptoms for Major Depressive Disorder (MDD). It consists of questions that are answered on a scale of 0 (not at all) to 3 (nearly every day). Previous studies have indicated that the PHQ-9 is a consistent and valid measure of depression severity~\cite{kroencke2001phq}. 

The GDS is a dichotomous ''yes'' or ''no'' evaluation tool commonly used to measure depression. While this scale has a long and short form, the long form of 30 questions was used to obtain most accurate and comprehensive results. This scale has been specifically tested and used extensively with older adults aged 65 and higher. Data shows that the GDS is reliable and promising in screening for depression in older adults~\cite{yesavage19869}. The results of our study are given in Table~\ref{tbl:phq9}. As the table shows 7 out of 10 participants had an improvement between 1 and 16 in their GDS depression score (the maximum score is 30) or between 1 and 6 in the PHQ-9 assessment (the maximum score is 27).

% Please add the following required packages to your document preamble:
% \usepackage{multirow}
\begin{table*}[t]
\caption{The exit survey evaluating Ryan's likeability and also sympathy with participants. The survey uses a five-point Likert scale.}
\label{tbl:survey}
\begin{tabular}{p{1cm}|p{\textwidth-5cm}|p{1cm}}
\cline{2-3}
 & Question & \multicolumn{1}{l|}{mean$\pm$std} \\ \hline
\multicolumn{1}{|l|}{\multirow{10}{*}{\STAB{\rotatebox[origin=c]{90}{ \specialcell{Evaluation of \\Ryan's Empathy \\and Emotion}}}}} & Q1. I felt Ryan was gentle with me. & \multicolumn{1}{l|}{4.80$\pm$.40} \\ \cline{2-3} 
\multicolumn{1}{|l|}{} & Q3. On scale of 1-5 how would you rate Ryan's facial expressions? & \multicolumn{1}{l|}{4.38$\pm$.70} \\ \cline{2-3} 
\multicolumn{1}{|l|}{} & Q4. I feel happier when I was in the company of Ryan. & \multicolumn{1}{l|}{4.50$\pm$.67} \\ \cline{2-3} 
\multicolumn{1}{|l|}{} & Q5. I feel less depressed after talking to Ryan. & \multicolumn{1}{l|}{4.50$\pm$.81} \\ \cline{2-3} 
\multicolumn{1}{|l|}{} & Q6. I felt Ryan understood my emotions. & \multicolumn{1}{l|}{4.10$\pm$1.04} \\ \cline{2-3} 
\multicolumn{1}{|l|}{} & Q7. Ryan encouraged me to open up about my mood/feelings. & \multicolumn{1}{l|}{4.00$\pm$1.18} \\ \cline{2-3} 
\multicolumn{1}{|l|}{} & Q8. The sessions with Ryan improved my mood and made me feel happier than I was before the session. & \multicolumn{1}{l|}{4.50$\pm$.92} \\ \cline{2-3} 
\multicolumn{1}{|l|}{} & Q10. How much do you agree with the statement: ``I like that Ryan empathized with me'' & \multicolumn{1}{l|}{4.57$\pm$.49} \\ \cline{2-3} 
\multicolumn{1}{|l|}{} & Q11. How well did Ryan empathize with your feelings? & \multicolumn{1}{l|}{4.40$\pm$.80} \\ \cline{2-3} 
\multicolumn{1}{|l|}{} & Q23. I feel happier when I was in the company of Ryan. & \multicolumn{1}{l|}{4.60$\pm$ .66} \\ \hline
\hline
%\multicolumn{3}{|l|}{} \\ \hline
\multicolumn{1}{|l|}{\multirow{18}{*}{\STAB{\rotatebox[origin=c]{90}{\specialcell{Evaluation of the interaction with \\Ryan and Likeability (Entire study)}}}}} & Q14. The conversation with Ryan was enjoyable. & \multicolumn{1}{l|}{4.50$\pm$.67} \\ \cline{2-3} 
\multicolumn{1}{|l|}{} & Q15. The conversation with Ryan was engaging. & \multicolumn{1}{l|}{4.20$\pm$.75} \\ \cline{2-3} 
\multicolumn{1}{|l|}{} & Q16. Learning to interact with Ryan was easy. & \multicolumn{1}{l|}{4.00$\pm$1.00} \\ \cline{2-3} 
\multicolumn{1}{|l|}{} & Q17. Talking with Ryan was like talking to a person. & \multicolumn{1}{l|}{3.90$\pm$1.37} \\ \cline{2-3} 
\multicolumn{1}{|l|}{} & Q18. I felt Ryan understood what I was saying. & \multicolumn{1}{l|}{4.00$\pm$1.26} \\ \cline{2-3} 
\multicolumn{1}{|l|}{} & Q19. Ryan was friendly. & \multicolumn{1}{l|}{4.80$\pm$.40} \\ \cline{2-3} 
\multicolumn{1}{|l|}{} & Q20. Ryan was likeable. & \multicolumn{1}{l|}{4.90$\pm$.30} \\ \cline{2-3} 
\multicolumn{1}{|l|}{} & Q21. Ryan was warm. & \multicolumn{1}{l|}{4.40$\pm$1.28} \\ \cline{2-3} 
\multicolumn{1}{|l|}{} & Q22. Ryan was intelligent. & \multicolumn{1}{l|}{4.70$\pm$.64} \\ \cline{2-3} 
\multicolumn{1}{|l|}{} & Q24. Ryan was acting natural. & \multicolumn{1}{l|}{4.50$\pm$.67} \\ \cline{2-3} 
\multicolumn{1}{|l|}{} & Q25. I would like to interact with Ryan again. & \multicolumn{1}{l|}{4.60$\pm$.66} \\ \cline{2-3} 
\multicolumn{1}{|l|}{} & Q26. I enjoyed interacting with Ryan at the end of week 3 as much as I did in the beginning of the study. & \multicolumn{1}{l|}{4.70$\pm$.46} \\ \cline{2-3} 
\multicolumn{1}{|l|}{} & Q27. I found myself looking forward to my sessions with Ryan. & \multicolumn{1}{l|}{4.70$\pm$.64} \\ \cline{2-3} 
\multicolumn{1}{|l|}{} & Q28. I enjoyed Ryan showing photos to me. & \multicolumn{1}{l|}{4.90$\pm$.30} \\ \cline{2-3} 
\multicolumn{1}{|l|}{} & Q29. I enjoyed Ryan playing videos for me. & \multicolumn{1}{l|}{4.90$\pm$.30} \\ \cline{2-3} 
\multicolumn{1}{|l|}{} & Q30. The videos played by Ryan were effective and helped me either learn something new or have a fun conversation. & \multicolumn{1}{l|}{4.90$\pm$.30} \\ \cline{2-3} 
\multicolumn{1}{|l|}{} & Q31. The conversations were organized and made sense. & \multicolumn{1}{l|}{4.80$\pm$.60} \\ \cline{2-3} 
\multicolumn{1}{|l|}{} & Q32. If given the change, I would continue further sessions with Ryan. & \multicolumn{1}{l|}{4.70$\pm$.64} \\ \hline
\hline
\multicolumn{1}{|l|}{\multirow{3}{*}{\STAB{\rotatebox[origin=c]{90}{Other Questions}}}} & \specialcell{Q12. We showed you two versions of Ryan, one with smile and empathy, and one without.\\Which versions of Ryan do you like the most?} & \multicolumn{1}{l|}{\specialcell{100\% selected the\\version with the\\smile expression}} \\ \cline{2-3} 
\multicolumn{1}{|l|}{} & \specialcell{ Q2/Q9:  did you notice a change in the way Ryan communicates with you and her \\ability in showing facial expressions after the session three crossover?} & \multicolumn{1}{l|}{73\% said yes.} \\ \cline{2-3} 
\multicolumn{1}{|l|}{} & \begin{tabular}[c]{@{}l@{}}\specialcell{Q33. On the scale of 1-5, how did you like the topics of the conversations Ryan had with you? \\ Response: Kids:4.56, Pets:3.75, TVShows: 3.5 Science:4.50, Music:4.50, \\Nature:4.88 Foods:4.38, Travel:4.63, Art:3.86, Movies:3.33 Reading:4.13, Sports: 3.44}\end{tabular} & \multicolumn{1}{l|}{} \\ \hline
\end{tabular}
\end{table*}
\subsection{Qualitative analysis}
\subsubsection{Exit survey questionnaires}
At the end of the study, we asked each participant to complete an exit survey. The survey contains 33 questions about the experiences they had with Ryan as follows: evaluation of Ryan's empathy and emotion and evaluation of the interaction with Ryan and the likeability of the conversation with Ryan and the conversation topics. We also asked the users to give us feedback about any other aspects of the robot and the study. The majority of questions were based on a five-point Likert scale where 1 means ``Strongly Disagree'' and 5 means ``Strongly Agree'', with an additional 5 ``yes'', ``no'' questions. Table~\ref{tbl:survey} reports the questions and the average score. It also shows the score for each topic.

Average score was above 4.00 on all the questions except question ``Q17: Talking with Ryan was like talking to a person'', where the average score was 3.90 (STD = 1.37). Notably, they gave an average score of 4.5 (STD = .67) on ``Q4: I feel happier when I was in the company of Ryan.'' and 4.57 (STD=.49) on ``Q10: How much do you agree that Ryan empathized with you''. We specifically asked participants ``Q2/Q9:  whether they noticed a change in the way Ryan communicates with them and its ability in showing facial expressions after the session three crossovers'' and 73\% of them said they noticed the change. 

\subsubsection{Participants' feedback}
In our exit interviews, we asked the participants to give us comments about the study and provide feedback on the experience they had with Ryan. The participants use the pronoun ``she/her'' to refer to Ryan since Ryan had a female face/voice in this study.  In the following we report the comments:

SN01: ``I had a good time. I enjoyed her very much. You want her to be a real thing like an addition to your home. I didn't think of her as a person like a dog or a cat.''

SN02: `` Ryan told me a lot of good things and I had a good time with her. She was very interesting and helpful.''

SN03: ``I liked her (``Ryan''). She is witty. At first, I didn't know what to think. I got better as I went. She sure has a pretty smile. It tears me up when she smiles, blinks her eyes. I would like to take her out to dinner but she wasn't hungry. Maybe next time.''

SN04: ``I liked her when she smiled. She interrupted me sometimes. Give me a chance to finish what I am saying. She was fun to talk to. I think the first one talked more I like with a smile. Very friendly.'' (Note: She is on G2 where Emotion was ON first and Ryan Smile and empathized).

\begin{table}[t]
\caption{Change in GDS and PHQ-9 Scores after participants completed the study. A negative (-) change means lower depression.}
\label{tbl:phq9}
\begin{tabular}{l|l|c|c|c|c|}
\cline{2-6}
\multirow{2}{*}{} & \multirow{2}{*}{\textbf{Sbj\#}} & \multicolumn{2}{c|}{\textbf{GDS*}} & \multicolumn{2}{c|}{\textbf{PHQ9*}} \\ \cline{3-6} 
 &  & \multicolumn{1}{c|}{Baseline} & \multicolumn{1}{c|}{\begin{tabular}[c]{@{}c@{}}Post-Study\\ Change\end{tabular}} & \multicolumn{1}{c|}{Baseline} & \multicolumn{1}{c|}{\begin{tabular}[c]{@{}c@{}}Post-Study\\ Change\end{tabular}} \\ \hline
\multicolumn{1}{|l|}{\multirow{5}{*}{  \STAB{\rotatebox[origin=c]{90}{\textbf{Group 1}}} }} & SN01 & 6/30 & +2 & 9/27 & +3 \\ \cline{2-6} 
\multicolumn{1}{|l|}{} & SN02 & 12/30 & -1 & 16/27 & -3 \\ \cline{2-6} 
\multicolumn{1}{|l|}{} & SN03 & 3/30 & -1 & 4/27 & -1 \\ \cline{2-6} 
\multicolumn{1}{|l|}{} & SN04 & 6/30 & +5 & 6/27 & -1 \\ \cline{2-6} 
\multicolumn{1}{|l|}{} & SN05 & 3/30 & -2 & 4/27 & -4 \\ \hline
\multicolumn{6}{|l|}{} \\ \hline
\multicolumn{1}{|l|}{\multirow{5}{*}{\STAB{\rotatebox[origin=c]{90}{\textbf{Group 2}}} }} & SN06 & 18/30 & -16 & 10/27 & -5 \\ \cline{2-6} 
\multicolumn{1}{|l|}{} & SN07 & 10/30 & -4 & 7/27 & -2 \\ \cline{2-6} 
\multicolumn{1}{|l|}{} & SN08 & 10/30 & -1 & 12/27 & -6 \\ \cline{2-6} 
\multicolumn{1}{|l|}{} & SN09 & 13/30 & -1 & 6/27 & +3 \\ \cline{2-6} 
\multicolumn{1}{|l|}{} & SN10 & 8/30 & 0 & 5/27 & +1 \\ \hline
\multicolumn{6}{|p{5cm}|}{\begin{tabular}[c]{p{\columnwidth-2cm}}*GDS: Normal: 0-9; mild depression: 10-19; Severe depression: 20-30. *PHQ9: Minimal Depression-0-4;Mild Depression-5-9; Moderate Depression-10-14;Moderately Sever Depression-15-19:Sever Depression-20-27\end{tabular}} \\ \hline
\end{tabular}
\end{table}

SN05: ``She was sort of creepy looking a little bit but she was fine. I was surprised I enjoyed it! I like her when she smiled. When she wasn't smiling she was kind of crummy.''

SN06: ``They forgot the eyelashes. The only thing I had difficulty was the lights. Took getting used to it.  I had so much fun in those meetings. Also, the thing was that when robot communicated and I paused, it would repeat itself.''

SN07: ``Enjoyed talking to Ryan. I would talk to her all the time if she was in my room. Good company. She needs eyelashes and smiling longer. The lights on the shoulder were sometimes frustrating. It would have been easier if it was just green.''

SN08: ``Ryan was very interesting and informative. When you first told me I was going to talk to a robot, I thought you were out of your mind but I really enjoyed it. She gave me ideas and information I had no ideas on.''

SN09: ``The longer I made an effort to communicate with Ryan the better it seemed to go. At a point, it became more natural to speak with the robot. She was cathartic.''

SN10: ``The robot asked a lot of questions and I didn't get to ask many questions. She looked really good. Her eyes blinked, her mouth moved. She smiled.''

\subsubsection{Caregiver's feedback}
We asked the participants' caregiver (staff member in  Eaton Senior Communities) about her observations of the subjects' behavior and mood pre- and post-study. Although the caregiver's observations are anecdotal and only represent one person's views/observations of subjects, it is still worth reviewing them since the caregiver had seen the subjects pre- and post-study and can judge changes in their well-being as an outsider.

She reported that the subjects who struggled with depression and social isolation benefited the most from interacting and conversing with Ryan. For instance, SN02 struggled with depression and social isolation (i.e. not attending holiday activities or no longer taking meals in the dining room), smiled and laughed again post-study and engaged in the community.  

The caregiver also reports that participants keep talking to her about the variations in Ryan's facial expression and particularly smiling as a feature that positively affected their relationship with Ryan. She reports that the improvement in mood was quickly apparent but also cognition, as residents were exposed to research and educational opportunities and ``stimulated human interaction.''

\section{Conclusion}
\label{sec:conclusion}

The growth of the elderly population and the widespread understaffing across nursing homes can exacerbate feelings of loneliness in the residents and overburden their nurses. During the COVID-19 pandemic, this issue became more evident~\cite{xu2020shortages}. The development of AI technologies drew attention to service robots and SAR as potential solutions to these problems. Robots may effectively relieve the burden on healthcare workers and improve the well-being of elderly individuals. Such robots need to be socio-emotionally intelligent in order to effectively engage the aging population. 

In this paper, we discussed Ryan, a socially assistive robot, and its multimodal emotion recognition and multimodal emotion expression systems. More specifically, we compared two versions of the robot: one that uses a scripted dialogue that does not factor in the users' emotions and is lacking facial expressions (non-empathic version), and one with facial expressions that uses an affective dialogue manager to generate a response and has the capability to recognize users' emotions (empathic version). 

We studied the differences and effects of Ryan's two versions with a cohort of older adults living in a senior care facility. The statistical analysis of the users' face-scale mood measurement (illustrated in Figure~\ref{fig:facescale}) indicates an overall positive effect as a result of the interaction with Ryan, irrespective of the robot being empathic or non-empathic. However, the word count measurement (Table~\ref{table:lmmresults}) and the exit survey analyses (Table~\ref{tbl:survey}) suggest that the empathic Ryan is perceived as more engaging and likable. Considering that the only difference between Ryan's two versions is empathic versus non-empathic, the findings suggest that empathy can encourage users to have longer conversations. Nonetheless, more experiments are needed to further study interactions using a more natural dialogue manager (chatbot). The changes in users' depression measurement scores (Table~\ref{tbl:phq9}) suggest that Ryan can potentially decrease users' depression, although to verify this finding more subjects and long-term studies are required. 

\section{Limitations and Future work}
\label{sec:futurework}
Although the study's results are positive and encouraging, our work has several limitations. Addressing these limitations in the future can improve Ryan and the effectiveness of similar HRI systems. When it comes to Ryan's perception and sensory input, acoustic signals and other modalities such as eye movement, gaze, head and body gesture, posture, and even breathing rhythm can be used to determine users' emotional state. Currently, Ryan does not utilize these sensory inputs, and adding these features would make the recognition of users' emotional state and intention more accurate and reliable. The other limitations of our study include small sample size, imbalanced participants' demographics, and the number of sessions.

Automatic speech recognition (ASR) is not specifically designed for older adults; their slow pace in talking and long pauses are considered ``End of Sentence''. This issue triggered the robot to interrupt participants, which had a negative effect on their perception. Open-domain dialogue is still an open question in computer science and consequently was the area that proved to have the most limitations in our study. While rule-based chatbots will never be perfect, our system still has room to grow in terms of the size of our knowledge base and our pattern matching rules. Finally, in this study, Ryan only mirrors the user's facial expression; in the future, the conversation and the context could drive the expression on Ryan's face.
 
\section{Acknowledgement}
Research reported in this manuscript was supported by the National Institute on Aging of the National Institutes of Health under award number R44AG059483 to DreamFace Technologlies, LLC. The content is solely the responsibility of the authors and does not necessarily represent the official views of the National Institutes of Health. The authors would like to thank Chandler Yunker and Gabriela Nordman for their contribution and help with dialogue writing.

\bibliographystyle{IEEEtran}
% argument is your BibTeX string definitions and bibliography database(s)
\bibliography{IEEEabrv,mybibfile}

% Generated by IEEEtran.bst, version: 1.14 (2015/08/26)
\begin{thebibliography}{10}
\providecommand{\url}[1]{#1}
\csname url@samestyle\endcsname
\providecommand{\newblock}{\relax}
\providecommand{\bibinfo}[2]{#2}
\providecommand{\BIBentrySTDinterwordspacing}{\spaceskip=0pt\relax}
\providecommand{\BIBentryALTinterwordstretchfactor}{4}
\providecommand{\BIBentryALTinterwordspacing}{\spaceskip=\fontdimen2\font plus
\BIBentryALTinterwordstretchfactor\fontdimen3\font minus
  \fontdimen4\font\relax}
\providecommand{\BIBforeignlanguage}[2]{{%
\expandafter\ifx\csname l@#1\endcsname\relax
\typeout{** WARNING: IEEEtran.bst: No hyphenation pattern has been}%
\typeout{** loaded for the language `#1'. Using the pattern for}%
\typeout{** the default language instead.}%
\else
\language=\csname l@#1\endcsname
\fi
#2}}
\providecommand{\BIBdecl}{\relax}
\BIBdecl

\bibitem{feil2005defining}
D.~Feil-Seifer and M.~J. Mataric, ``Defining socially assistive robotics,'' in
  \emph{Rehabilitation Robotics, 2005. ICORR 2005. 9th International Conference
  on}.\hskip 1em plus 0.5em minus 0.4em\relax IEEE, 2005, pp. 465--468.

\bibitem{banerjee2020social}
D.~Banerjee, M.~Rai \emph{et~al.}, ``Social isolation in covid-19: The impact
  of loneliness,'' \emph{International Journal of Social Psychiatry}, vol.~66,
  no.~6, pp. 525--527, 2020.

\bibitem{aung2017loneliness}
K.~Aung, M.~S. Nurumal, and W.~Bukhari, ``Loneliness among elderly in nursing
  homes,'' \emph{International Journal for Studies on Children, Women, Elderly
  And Disabled}, vol.~2, pp. 72--8, 2017.

\bibitem{ghafurian2021social}
M.~Ghafurian, C.~Ellard, and K.~Dautenhahn, ``Social companion robots to reduce
  isolation: a perception change due to covid-19,'' in \emph{IFIP Conference on
  Human-Computer Interaction}.\hskip 1em plus 0.5em minus 0.4em\relax Springer,
  2021, pp. 43--63.

\bibitem{vandemeulebroucke2018older}
T.~Vandemeulebroucke, B.~D. de~Casterl{\'e}, and C.~Gastmans, ``How do older
  adults experience and perceive socially assistive robots in aged care: a
  systematic review of qualitative evidence,'' \emph{Aging \& mental health},
  vol.~22, no.~2, pp. 149--167, 2018.

\bibitem{xu2020shortages}
H.~Xu, O.~Intrator, and J.~R. Bowblis, ``Shortages of staff in nursing homes
  during the covid-19 pandemic: What are the driving factors?'' \emph{Journal
  of the American Medical Directors Association}, vol.~21, no.~10, pp.
  1371--1377, 2020.

\bibitem{chen2020social}
S.-C. Chen, W.~Moyle, C.~Jones, and H.~Petsky, ``A social robot intervention on
  depression, loneliness, and quality of life for taiwanese older adults in
  long-term care,'' \emph{International psychogeriatrics}, vol.~32, no.~8, pp.
  981--991, 2020.

\bibitem{adams_social_2020}
\BIBentryALTinterwordspacing
A.~Adams, J.~Beer, X.~Wu, J.~Komsky, and J.~Zamer, ``Social {Activities} in
  {Community} {Settings}: {Impact} of {COVID}-19 and {Technology}
  {Solutions},'' \emph{Innovation in Aging}, vol.~4, no. Suppl 1, p. 957, Dec.
  2020. [Online]. Available:
  \url{https://www.ncbi.nlm.nih.gov/pmc/articles/PMC7741293/}
\BIBentrySTDinterwordspacing

\bibitem{henkel2020robotic}
A.~P. Henkel, M.~{\v{C}}ai{\'c}, M.~Blaurock, and M.~Okan, ``Robotic
  transformative service research: deploying social robots for consumer
  well-being during covid-19 and beyond,'' \emph{Journal of Service
  Management}, 2020.

\bibitem{tapus2007grand}
A.~Tapus, M.~Maja, and B.~Scassellatti, ``The grand challenges in socially
  assistive robotics,'' \emph{IEEE Robotics and Automation Magazine}, vol.~14,
  no.~1, pp. 35--42, 2007.

\bibitem{prendinger2005empathic}
H.~Prendinger and M.~Ishizuka, ``The empathic companion: A character-based
  interface that addresses users'affective states,'' \emph{Applied Artificial
  Intelligence}, vol.~19, no. 3-4, pp. 267--285, 2005.

\bibitem{bagheri2021reinforcement}
E.~Bagheri, O.~Roesler, H.-L. Cao, and B.~Vanderborght, ``A reinforcement
  learning based cognitive empathy framework for social robots,''
  \emph{International Journal of Social Robotics}, vol.~13, no.~5, pp.
  1079--1093, 2021.

\bibitem{yonck2020heart}
R.~Yonck, \emph{Heart of the machine: Our future in a world of artificial
  emotional intelligence}.\hskip 1em plus 0.5em minus 0.4em\relax Arcade, 2020.

\bibitem{pu2019effectiveness}
L.~Pu, W.~Moyle, C.~Jones, and M.~Todorovic, ``The effectiveness of social
  robots for older adults: a systematic review and meta-analysis of randomized
  controlled studies,'' \emph{The Gerontologist}, vol.~59, no.~1, pp. e37--e51,
  2019.

\bibitem{Nielsen2010APF}
R.~Nielsen, R.~Voyles, D.~Bola{\~n}os, M.~Mahoor, W.~Pace, K.~Siek, and
  W.~Ward, ``A platform for human-robot dialog systems research,'' in
  \emph{AAAI Fall Symposium: Dialog with Robots}, 2010, pp. 161--162.

\bibitem{picard2000affective}
R.~W. Picard, \emph{Affective computing}.\hskip 1em plus 0.5em minus
  0.4em\relax MIT press, 2000.

\bibitem{10.3389/fphar.2020.00279}
\BIBentryALTinterwordspacing
C.~Linnemann and U.~E. Lang, ``Pathways connecting late-life depression and
  dementia,'' \emph{Frontiers in Pharmacology}, vol.~11, p. 279, 2020.
  [Online]. Available:
  \url{https://www.frontiersin.org/article/10.3389/fphar.2020.00279}
\BIBentrySTDinterwordspacing

\bibitem{abdollahi2017pilot}
H.~Abdollahi, A.~Mollahosseini, J.~T. Lane, and M.~Mahoor, ``A pilot study on
  using an intelligent life-like robot as a companion for elderly individuals
  with dementia and depression,'' in \emph{2017 IEEE-RAS 17th International
  Conference on Humanoid Robotics (Humanoids)}, Nov 2017, pp. 541--546.

\bibitem{Brackett2011}
M.~A. Brackett, S.~E. Rivers, and P.~Salovey, ``{Emotional intelligence:
  Implications for personal, social, academic, and workplace success},''
  \emph{Social and Personality Psychology Compass}, vol.~5, 2011.

\bibitem{ochs2005intelligent}
M.~Ochs, R.~Niewiadomski, C.~Pelachaud, and D.~Sadek, ``Intelligent expressions
  of emotions,'' in \emph{International Conference on Affective Computing and
  Intelligent Interaction}.\hskip 1em plus 0.5em minus 0.4em\relax Springer,
  2005, pp. 707--714.

\bibitem{mcduff2018designing}
D.~McDuff and M.~Czerwinski, ``Designing emotionally sentient agents,''
  \emph{Communications of the ACM}, vol.~61, no.~12, pp. 74--83, 2018.

\bibitem{alvarez2010emotional}
M.~{\'A}lvarez, R.~Gal{\'a}n, F.~Mat{\'\i}a, D.~Rodr{\'\i}guez-Losada, and
  A.~Jim{\'e}nez, ``An emotional model for a guide robot,'' \emph{IEEE
  Transactions on Systems, Man, and Cybernetics-Part A: Systems and Humans},
  vol.~40, no.~5, pp. 982--992, 2010.

\bibitem{preston2002empathy}
S.~D. Preston and F.~B. De~Waal, ``Empathy: Its ultimate and proximate bases,''
  \emph{Behavioral and brain sciences}, vol.~25, no.~1, pp. 1--20, 2002.

\bibitem{mollahosseini2018studying}
A.~Mollahosseini, H.~Abdollahi, and M.~H. Mahoor, ``Studying effects of
  incorporating automated affect perception with spoken dialog in social
  robots,'' in \emph{2018 27th IEEE International Symposium on Robot and Human
  Interactive Communication (RO-MAN)}.\hskip 1em plus 0.5em minus 0.4em\relax
  IEEE, 2018, pp. 783--789.

\bibitem{paiva2005learning}
A.~Paiva, J.~Dias, D.~Sobral, R.~Aylett, S.~Woods, L.~Hall, and C.~Zoll,
  ``Learning by feeling: Evoking empathy with synthetic characters,''
  \emph{Applied Artificial Intelligence}, vol.~19, no. 3-4, pp. 235--266, 2005.

\bibitem{leite2013influence}
I.~Leite, A.~Pereira, S.~Mascarenhas, C.~Martinho, R.~Prada, and A.~Paiva,
  ``The influence of empathy in human--robot relations,'' \emph{International
  journal of human-computer studies}, vol.~71, no.~3, pp. 250--260, 2013.

\bibitem{alves2019empathic}
P.~Alves-Oliveira, P.~Sequeira, F.~S. Melo, G.~Castellano, and A.~Paiva,
  ``Empathic robot for group learning: A field study,'' \emph{ACM Transactions
  on Human-Robot Interaction (THRI)}, vol.~8, no.~1, pp. 1--34, 2019.

\bibitem{castellano2008emotion}
G.~Castellano, L.~Kessous, and G.~Caridakis, ``Emotion recognition through
  multiple modalities: face, body gesture, speech,'' in \emph{Affect and
  emotion in human-computer interaction}.\hskip 1em plus 0.5em minus
  0.4em\relax Springer, 2008, pp. 92--103.

\bibitem{spezialetti2020emotion}
M.~Spezialetti, G.~Placidi, and S.~Rossi, ``Emotion recognition for human-robot
  interaction: recent advances and future perspectives,'' \emph{Frontiers in
  Robotics and AI}, vol.~7, 2020.

\bibitem{wada2003effects}
K.~Wada, T.~Shibata, T.~Saito, and K.~Tanie, ``Effects of robot assisted
  activity to elderly people who stay at a health service facility for the
  aged,'' in \emph{Intelligent Robots and Systems, 2003.(IROS 2003).
  Proceedings. 2003 IEEE/RSJ International Conference on}, vol.~3.\hskip 1em
  plus 0.5em minus 0.4em\relax IEEE, 2003, pp. 2847--2852.

\bibitem{petersen2017utilization}
S.~Petersen, S.~Houston, H.~Qin, C.~Tague, and J.~Studley, ``The utilization of
  robotic pets in dementia care,'' \emph{Journal of Alzheimer's Disease},
  vol.~55, no.~2, pp. 569--574, 2017.

\bibitem{Francesca2019}
F.~Dino, R.~Zandie, H.~Abdollahi, S.~Schoeder, and M.~Mahoor, ``Delivering
  cognitive behavioral therapy using a conversational social robot,'' in
  \emph{Intelligent Robots and Systems (IROS), IEEE/RSJ International
  Conference on}, 2019.

\bibitem{sarabia2018assistive}
M.~Sarabia, N.~Young, K.~Canavan, T.~Edginton, Y.~Demiris, and M.~P.
  Vizcaychipi, ``Assistive robotic technology to combat social isolation in
  acute hospital settings,'' \emph{International Journal of Social Robotics},
  vol.~10, no.~5, pp. 607--620, 2018.

\bibitem{nao}
``Nao,'' \url{https://www.softbankrobotics.com/emea/en/robots/nao}, accessed:
  2018-09-13.

\bibitem{brackett2011emotional}
M.~A. Brackett, S.~E. Rivers, and P.~Salovey, ``Emotional intelligence:
  Implications for personal, social, academic, and workplace success,''
  \emph{Social and Personality Psychology Compass}, vol.~5, no.~1, pp. 88--103,
  2011.

\bibitem{pantic2005affective}
M.~Pantic, N.~Sebe, J.~F. Cohn, and T.~Huang, ``Affective multimodal
  human-computer interaction,'' in \emph{Proceedings of the 13th annual ACM
  international conference on Multimedia}.\hskip 1em plus 0.5em minus
  0.4em\relax ACM, 2005, pp. 669--676.

\bibitem{sebe2005multimodal}
N.~Sebe, I.~Cohen, and T.~S. Huang, ``Multimodal emotion recognition,'' in
  \emph{Handbook of Pattern Recognition and Computer Vision}.\hskip 1em plus
  0.5em minus 0.4em\relax World Scientific, 2005, pp. 387--409.

\bibitem{shi2018sentiment}
W.~Shi and Z.~Yu, ``Sentiment adaptive end-to-end dialog systems,'' in
  \emph{Proceedings of the 56th Annual Meeting of the Association for
  Computational Linguistics (Volume 1: Long Papers)}, 2018, pp. 1509--1519.

\bibitem{cowie1996automatic}
R.~Cowie and E.~Douglas-Cowie, ``Automatic statistical analysis of the signal
  and prosodic signs of emotion in speech,'' in \emph{Proceeding of Fourth
  International Conference on Spoken Language Processing. ICSLP'96},
  vol.~3.\hskip 1em plus 0.5em minus 0.4em\relax IEEE, 1996, pp. 1989--1992.

\bibitem{bianchi2003categorical}
N.~Bianchi-Berthouze and A.~Kleinsmith, ``A categorical approach to affective
  gesture recognition,'' \emph{Connection science}, vol.~15, no.~4, pp.
  259--269, 2003.

\bibitem{dapretto2006understanding}
M.~Dapretto, M.~S. Davies, J.~H. Pfeifer, A.~A. Scott, M.~Sigman, S.~Y.
  Bookheimer, and M.~Iacoboni, ``Understanding emotions in others: mirror
  neuron dysfunction in children with autism spectrum disorders,'' \emph{Nature
  neuroscience}, vol.~9, no.~1, p.~28, 2006.

\bibitem{hess2001facial}
U.~Hess and S.~Blairy, ``Facial mimicry and emotional contagion to dynamic
  emotional facial expressions and their influence on decoding accuracy,''
  \emph{International journal of psychophysiology}, vol.~40, no.~2, pp.
  129--141, 2001.

\bibitem{pepper}
SoftBank, ``Pepper robot,''
  \url{https://www.softbankrobotics.com/emea/en/pepper/}, 2021, [Online;
  accessed 19-July-2021].

\bibitem{dreamfacetech}
\BIBentryALTinterwordspacing
DreamFace-Tech., ``Social robotics,'' 2015, last checked: 01.20.2017. [Online].
  Available: \url{http://dreamfacetech.com/}
\BIBentrySTDinterwordspacing

\bibitem{zeno}
\BIBentryALTinterwordspacing
Zeno, ``Zeno,'' 2009. [Online]. Available:
  \url{http://www.hansonrobotics.com/robot/zeno/}
\BIBentrySTDinterwordspacing

\bibitem{socibot}
\BIBentryALTinterwordspacing
Socibot, ``Socibot platform of engineeredarts,'' 2015, last checked:
  01.20.2017. [Online]. Available:
  \url{https://www.engineeredarts.co.uk/socibot/}
\BIBentrySTDinterwordspacing

\bibitem{kanamori2003pilot}
M.~Kanamori, M.~Suzuki, H.~Oshiro, M.~Tanaka, T.~Inoguchi, H.~Takasugi,
  Y.~Saito, and T.~Yokoyama, ``Pilot study on improvement of quality of life
  among elderly using a pet-type robot,'' in \emph{Proceedings 2003 IEEE
  International Symposium on Computational Intelligence in Robotics and
  Automation. Computational Intelligence in Robotics and Automation for the New
  Millennium (Cat. No. 03EX694)}, vol.~1.\hskip 1em plus 0.5em minus
  0.4em\relax IEEE, 2003, pp. 107--112.

\bibitem{Kotwal_2016}
\BIBentryALTinterwordspacing
A.~A. Kotwal, J.~Kim, L.~Waite, and W.~Dale, ``Social function and cognitive
  status: Results from a us nationally representative survey of older adults,''
  \emph{Journal of General Internal Medicine}, vol.~31, no.~8, p. 854–862,
  Apr 2016. [Online]. Available:
  \url{http://dx.doi.org/10.1007/s11606-016-3696-0}
\BIBentrySTDinterwordspacing

\bibitem{zamora2017association}
M.~Zamora-Macorra, E.~F.~A. de~Castro, J.~A. {\'A}vila-Funes, B.~S.
  Manrique-Espinoza, R.~L{\'o}pez-Ridaura, A.~L. Sosa-Ortiz, P.~L. Shields, and
  D.~S.~M. del Campo, ``The association between social support and cognitive
  function in mexican adults aged 50 and older,'' \emph{Archives of Gerontology
  and Geriatrics}, vol.~68, pp. 113--118, 2017.

\bibitem{sander2013models}
D.~Sander, ``Models of emotion: The affective neuroscience approach.''
  \emph{The Cambridge Handbook of Human Affective Neuroscience}, 2013.

\bibitem{russell1980circumplex}
J.~Russell, ``{A circumplex model of affect},'' \emph{Journal of personality
  and social psychology}, vol.~39, no.~6, pp. 1161--1178, 1980.

\bibitem{Ekman1978}
P.~Ekman and W.~Friesen, \emph{The Facial Action Coding System: A Technique for
  the Measurement of Facial Movement}.\hskip 1em plus 0.5em minus 0.4em\relax
  Consulting Psychologists Press, 1978.

\bibitem{szaboova2020emotion}
M.~Szab{\'o}ov{\'a}, M.~Sarnovsk{\`y},
  V.~Maslej~Kre{\v{s}}{\v{n}}{\'a}kov{\'a}, and K.~Machov{\'a}, ``Emotion
  analysis in human--robot interaction,'' \emph{Electronics}, vol.~9, no.~11,
  p. 1761, 2020.

\bibitem{cavallo2018emotion}
F.~Cavallo, F.~Semeraro, L.~Fiorini, G.~Magyar, P.~Sin{\v{c}}{\'a}k, and
  P.~Dario, ``Emotion modelling for social robotics applications: a review,''
  \emph{Journal of Bionic Engineering}, vol.~15, no.~2, pp. 185--203, 2018.

\bibitem{viola2004robust}
P.~Viola and M.~J. Jones, ``Robust real-time face detection,''
  \emph{International journal of computer vision}, vol.~57, no.~2, pp.
  137--154, 2004.

\bibitem{szegedy2015going}
C.~Szegedy, W.~Liu, Y.~Jia, P.~Sermanet, S.~Reed, D.~Anguelov, D.~Erhan,
  V.~Vanhoucke, and A.~Rabinovich, ``Going deeper with convolutions,'' in
  \emph{Proceedings of the IEEE Conference on Computer Vision and Pattern
  Recognition}, 2015, pp. 1--9.

\bibitem{mollahosseini2017affectnet}
A.~Mollahosseini, B.~Hasani, and M.~H. Mahoor, ``Affectnet: A database for
  facial expression, valence, and arousal computing in the wild,'' \emph{IEEE
  Transactions on Affective Computing}, vol.~PP, no.~99, pp. 1--1, 2017.

\bibitem{manning-EtAl:2014:P14-5}
\BIBentryALTinterwordspacing
C.~D. Manning, M.~Surdeanu, J.~Bauer, J.~Finkel, S.~J. Bethard, and
  D.~McClosky, ``The {Stanford} {CoreNLP} natural language processing
  toolkit,'' in \emph{Association for Computational Linguistics (ACL) System
  Demonstrations}, 2014, pp. 55--60. [Online]. Available:
  \url{http://www.aclweb.org/anthology/P/P14/P14-5010}
\BIBentrySTDinterwordspacing

\bibitem{program-y}
Keiffster, ``Program-y,'' \url{https://github.com/keiffster/program-y}, 2021,
  [Online; accessed 11-June-2021].

\bibitem{Burkhardt2009EmotionDI}
F.~Burkhardt, M.~V. Ballegooy, K.-P. Engelbrecht, T.~Polzehl, and J.~Stegmann,
  ``Emotion detection in dialog systems: Applications, strategies and
  challenges,'' \emph{2009 3rd International Conference on Affective Computing
  and Intelligent Interaction and Workshops}, pp. 1--6, 2009.

\bibitem{bertero2016real}
D.~Bertero, F.~B. Siddique, C.-S. Wu, Y.~Wan, R.~H.~Y. Chan, and P.~Fung,
  ``Real-time speech emotion and sentiment recognition for interactive dialogue
  systems,'' in \emph{Proceedings of the 2016 Conference on Empirical Methods
  in Natural Language Processing}, 2016, pp. 1042--1047.

\bibitem{Nwe2003SpeechER}
T.~L. Nwe, S.~W. Foo, and L.~C.~D. Silva, ``Speech emotion recognition using
  hidden markov models,'' \emph{Speech Communication}, vol.~41, pp. 603--623,
  2003.

\bibitem{wallace2003elements}
R.~Wallace, ``The elements of aiml style,'' \emph{Alice AI Foundation}, vol.
  139, 2003.

\bibitem{Wallace2009}
R.~S. Wallace, ``The anatomy of alice,'' in \emph{Parsing the Turing
  Test}.\hskip 1em plus 0.5em minus 0.4em\relax Springer, 2009, pp. 181--210.

\bibitem{restfulapi}
``Restful api,'' \url{https://restfulapi.net/}, accessed: 2018-09-13.

\bibitem{deng2019embodiment}
E.~Deng, B.~Mutlu, M.~J. Mataric \emph{et~al.}, ``Embodiment in socially
  interactive robots,'' \emph{Foundations and Trends{\textregistered} in
  Robotics}, vol.~7, no.~4, pp. 251--356, 2019.

\bibitem{romano2005basic}
D.~M. Romano, G.~Sheppard, J.~Hall, A.~Miller, and Z.~Ma, ``Basic: A
  believable, adaptable socially intelligent character for social presence,''
  in \emph{Proceedings of the 8th Annual International Workshop on
  Presence}.\hskip 1em plus 0.5em minus 0.4em\relax Citeseer, 2005, pp.
  287--290.

\bibitem{kasap2009making}
Z.~Kasap, M.~B. Moussa, P.~Chaudhuri, and N.~Magnenat-Thalmann, ``Making them
  remember—emotional virtual characters with memory,'' \emph{IEEE Computer
  Graphics and Applications}, vol.~29, no.~2, pp. 20--29, 2009.

\bibitem{schroder2011building}
M.~Schroder, E.~Bevacqua, R.~Cowie, F.~Eyben, H.~Gunes, D.~Heylen, M.~Ter~Maat,
  G.~McKeown, S.~Pammi, M.~Pantic \emph{et~al.}, ``Building autonomous
  sensitive artificial listeners,'' \emph{IEEE transactions on affective
  computing}, vol.~3, no.~2, pp. 165--183, 2011.

\bibitem{devault2014simsensei}
D.~DeVault, R.~Artstein, G.~Benn, T.~Dey, E.~Fast, A.~Gainer, K.~Georgila,
  J.~Gratch, A.~Hartholt, M.~Lhommet \emph{et~al.}, ``Simsensei kiosk: A
  virtual human interviewer for healthcare decision support,'' in
  \emph{Proceedings of the 2014 international conference on Autonomous agents
  and multi-agent systems}, 2014, pp. 1061--1068.

\bibitem{irfan2020dynamic}
B.~Irfan, A.~Narayanan, and J.~Kennedy, ``Dynamic emotional language adaptation
  in multiparty interactions with agents,'' in \emph{Proceedings of the 20th
  ACM International Conference on Intelligent Virtual Agents}, 2020, pp. 1--8.

\bibitem{mollahosseini2018role}
A.~Mollahosseini, H.~Abdollahi, T.~Sweeny, R.~Cole, and M.~Mahoor, ``Role of
  embodiment and presence in human perception of robots’ facial cues,''
  \emph{International Journal of Human-Computer Studies}, vol. 116, pp. 25--39,
  2018.

\bibitem{tariq2006saint}
S.~Tariq, N.~Tumosa, J.~Chibnall, H.~Perry~III, and J.~Morley, ``The saint
  louis university mental status (slums) examination for detecting mild
  cognitive impairment and dementia is more sensitive than the mini-mental
  status examination (mmse)--a pilot study,'' \emph{Am J Geriatr Psychiatry},
  vol.~14, no.~11, pp. 900--10, 2006.

\bibitem{hill2015real}
J.~Hill, W.~R. Ford, and I.~G. Farreras, ``Real conversations with artificial
  intelligence: A comparison between human--human online conversations and
  human--chatbot conversations,'' \emph{Computers in human behavior}, vol.~49,
  pp. 245--250, 2015.

\bibitem{lorish1986face}
C.~D. Lorish and R.~Maisiak, ``The face scale: a brief, nonverbal method for
  assessing patient mood,'' \emph{Arthritis \& Rheumatism: Official Journal of
  the American College of Rheumatology}, vol.~29, no.~7, pp. 906--909, 1986.

\bibitem{kargar2017pilot}
A.~H. Kargar~B and M.~H. Mahoor, ``A {Pilot} {Study} on the {eBear} {Socially}
  {Assistive} {Robot}: {Implication} for {Interacting} with {Elderly} {People}
  with {Moderate} {Depression},'' in \emph{{IEEE}-{RAS} {International}
  {Conference} on {Humanoid} {Robots}}, Birmingham, UK, Nov. 2017.

\bibitem{EmoteProject}
\BIBentryALTinterwordspacing
T.~E. Project, ``The emote project,'' 2013, last checked: 01.20.2017. [Online].
  Available: \url{http://gaips.inesc-id.pt/emote/}
\BIBentrySTDinterwordspacing

\bibitem{davis1983measuring}
M.~H. Davis, ``Measuring individual differences in empathy: Evidence for a
  multidimensional approach.'' \emph{Journal of personality and social
  psychology}, vol.~44, no.~1, p. 113, 1983.

\bibitem{woolson2007wilcoxon}
R.~F. Woolson, ``Wilcoxon signed-rank test,'' \emph{Wiley encyclopedia of
  clinical trials}, pp. 1--3, 2007.

\bibitem{kroencke2001phq}
K.~Kroencke, R.~Spitzer, and J.~Williams, ``The phq-9: validity of a brief
  depression severity measure [electronic version],'' \emph{Journal of General
  Internal Medicine}, vol.~16, no.~9, pp. 606--13, 2001.

\bibitem{yesavage19869}
J.~A. Yesavage and J.~I. Sheikh, ``9/geriatric depression scale (gds) recent
  evidence and development of a shorter version,'' \emph{Clinical
  gerontologist}, vol.~5, no. 1-2, pp. 165--173, 1986.

\end{thebibliography}

\vskip -2.0cm
\begin{IEEEbiography}[{\includegraphics[width=1in,height=1.25in,clip,keepaspectratio]{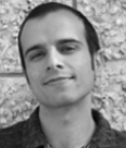}}]{Hojjat Abdollahi} received his Masters in Artificial Intelligence from Sharif University of Technology, Tehran, Iran. He is the VP of engineering at DreamFace Technologies, LLC and currently and also a Ph.D. Student of Electrical and Computer Engineering at the University of Denver, USA. During his Ph.D. program, Hojjat has focused on developing and studying a social robot called Ryan the Companionbot. His work at the University of Denver focuses specifically on Social Robotics and Human-Robot Interaction.
\end{IEEEbiography}

\vskip -2.0cm
\begin{IEEEbiography}[{\includegraphics[width=1in,height=1.25in,clip,keepaspectratio]{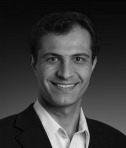}}]{Mohammad H. Mahoor} received the Ph.D. degree in electrical and computer engineering from the University of Miami, Florida, in 2007. He is a Professor of Electrical and Computer Engineering at DU and also the founder of DreamFace Technologies, LLC, a start-up robotics company aims at developing and commercializing Ryan companionbot for assisting elder adults with depression and dementia. He does research in the area of computer vision, affective computing, and human-robot interaction (HRI). His research focus is on designing humanoid social robots for assisting  children with special needs (i.e. autism) and older adults with depression and dementia. He is a Senior Member of IEEE and has published about 130 peer-reviewed conference and journal papers.
\end{IEEEbiography}

\vskip -2.0cm
\begin{IEEEbiography}[{\includegraphics[width=1in,height=1.25in,clip,keepaspectratio]{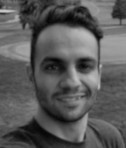}}]{Rohola Zandie} received the MSc degree in computer engineering - artificial intelligence from Sharif University of Technology, Tehran, Iran, in 2015. He has frequently worked as an intern  at DreamFace Technologies, LLC and while pursuing his Ph.D. degree in electrical and computer engineering and is a graduate research assistant in the Department of Electrical and Computer Engineering at the University of Denver. His research interests include Natural Language Processing and Deep Neural Networks.
\end{IEEEbiography}

\vskip -2.0cm
\begin{IEEEbiography}[{\includegraphics[width=1in,height=1.25in,clip,keepaspectratio]{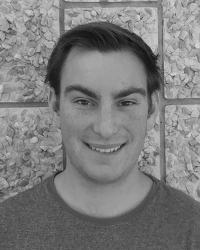}}]{Jarid Siewierski} received the BSc degree in computer science from the University of Denver in 2018. He has been with DreamFace Technologies, LLC since 2018 working on developing chatbots and dialogue management systems utilized in social robots. His research interests include Natural Language Processing, and dialogue systems.
\end{IEEEbiography}

\vskip -2.0cm
\begin{IEEEbiography}[{\includegraphics[width=1in,height=1.25in,clip,keepaspectratio]{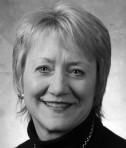}}]{Sara H. Qualls} is the Kraemer Family Professor of Aging Studies and Psychology, and director of the Gerontology Center at the University of Colorado at Colorado Springs (UCCS). She led the development of the doctoral program in Clinical Psychology (Emphasizing Geropsychology) at UCCS, and the establishment of the UCCS Aging Center which provides mental health and family support services for older adults in the region. Qualls coordinates efforts of UCCS faculty and students on behalf of the community in collaborative projects.  She has published widely on mental health and aging, including several books.  
\end{IEEEbiography}
% You can push biographies down or up by placing
% a \vfill before or after them. The appropriate
% use of \vfill depends on what kind of text is
% on the last page and whether or not the columns
% are being equalized.

% \vfill

% Can be used to pull up biographies so that the bottom of the last one
% is flush with the other column.
% \enlargethispage{-5in}

\end{document}